\documentclass[a4paper,11pt]{article}
\usepackage{jheppub} 
\usepackage{lineno}
\usepackage{mathtools}
\usepackage{mathrsfs}
\usepackage{bbm}
\usepackage{enumitem}
\usepackage{quiver}
\usepackage{tikz}
\usepackage{comment}
\usetikzlibrary{positioning}
\usepackage[dvipsnames]{xcolor}

\newcommand{\ket}[1]{\left|#1\right\rangle}
\newcommand{\bra}[1]{\left\langle#1\right|}
\newcommand{\ip}[2]{\left\langle#1\middle|#2\right\rangle}
\newcommand{\avg}[1]{\left\langle#1\right\rangle_{\mu}}
\newcommand{\avgn}[1]{\left\langle#1\right\rangle_{\mu_n}}
\newcommand{\wig}[3]{D^{#1}_{#2}(#3)}

\usepackage{color}

\title{Holographic Codes from Enriched Link Entanglement in Spin Networks
}

\author[a]{Mai Qi}
\author[a,b,c,d]{Eugenia Colafranceschi}

\affiliation[a]{Department of Physics, University of California, Santa Barbara, CA 93106, USA}
\affiliation[b]{Department of Physics and Astronomy, Western University, N6A 3K7 London, ON, Canada}
\affiliation[c]{Departamento de Física Teórica,
Universidad Complutense de Madrid,
Plaza de las Ciencias 1, 28040 Madrid, Spain}
\affiliation[d]{Department of Applied Mathematics, University of Waterloo, 200 University Avenue West, N2L 3G1 Waterloo, ON, Canada}

\emailAdd{maiqi@ucsb.edu, ecolafra@uwaterloo.ca}

\abstract{
We introduce an enriched entanglement structure for spin networks, inspired by tensor-network constructions, in which internal links can carry a controlled and discrete amount of entanglement. In the spin-network picture, vertices are dual to simplices (more generally, to polyhedra) and links are dual to their faces. Standard spin-network gluing corresponds to fully identifying two simplices along a face, implemented by a maximally entangled, gauge-invariant singlet state on the corresponding link, while unglued faces correspond to links carrying no entanglement. Working on a complete graph, we promote this binary choice to a controlled and tunable structure by allowing each link to carry a variable amount of entanglement, interpolating between product states and the fully entangled singlet. The additional link variables therefore control not only the amount of entanglement but also the extent to which gauge invariance at internal links is preserved or broken, admitting an interpretation in terms of emergent edge-mode--like degrees of freedom. Within this framework, spin-network contraction defines a bulk–to–boundary map from link-entanglement data to boundary states. Adapting techniques developed in random tensor networks, we show that in a suitable large-spin regime the map is a co-isometry in expectation value, meaning that its adjoint preserves boundary inner products on average, up to the overall scale. Restricting to a code subspace defined by configurations in which links are either effectively glued or open, with small fluctuations around this pattern, the map becomes an isometry up to corrections that vanish in the large-spin limit. This yields a discrete and geometrically meaningful realization of holographic and error-correcting features within the spin-network Hilbert space.
}

\begin{document}
\maketitle
\flushbottom

\newpage
\section{Introduction}

 The difficulty of formulating a consistent theory of quantum gravity, together with the current lack of direct experimental guidance, has led to the development of a wide range of theoretical approaches. While this diversity is a strength, it also raises the question of how results obtained in one approach can be meaningfully compared with, or translated into, another. A growing line of research therefore aims at building connections across different quantum gravity frameworks, with the goal of identifying shared structures and facilitating the transfer of ideas, tools, and results.\smallskip

Some approaches rely on continuum quantum field–theoretic tools, such as String Theory and AdS/CFT, the latter providing a concrete realization of the holographic principle: a $(d+1)$-dimensional spacetime with gravity is encoded in a $d$-dimensional quantum field theory without gravity. Others aim to reconstruct continuum spacetime from fundamentally discrete geometric structures, as in canonical Loop Quantum Gravity (LQG), Spin Foams, Causal Set Theory, Causal Dynamical Triangulations; a ``second-quantization version'' of LQG, Group Field Theory (GFT), occupies an intermediate position, combining quantum field–theoretic techniques with a description of quantum spacetime in terms of discrete building blocks. Bridging these perspectives is challenging, not only because of different views (e.g. on the role of the holographic principle), but also due to the markedly different mathematical and conceptual tools they employ.\smallskip

Yet intriguing connections between holography and discrete geometry arise in the context of tensor networks, tools originally devised to study many-body entanglement and renormalization in condensed-matter systems, which encode entanglement in a discrete and combinatorial structure. On the one hand, tensor-network constructions have provided useful toy models for holography: random tensor networks (RTN)~\cite{HoloRTN} reproduce area-law entanglement, admit an effective Ryu–Takayanagi (RT) prescription~\cite{Harlow:2016vwg}, and realize a form of entanglement-wedge reconstruction~\cite{Pastawski:2015qua}, whereby bulk operators supported in a given region can be mapped to boundary operators acting on a corresponding subsystem.\\On the other hand, tensor networks display striking analogies with spin networks, which provide a kinematical description of quantum geometry in canonical LQG, Spin Foams, GFT and related approaches. Spin networks are graphs dual to simplicial (more generally, polyhedral) decompositions of space: links are labelled by representations of a gauge group (typically $SU(2)$), encoding the areas of the dual faces, while nodes represent simplices and are labelled by gauge-invariant tensors (intertwiners), which encode the volume and shape degrees of freedom of the dual building blocks. \smallskip

A careful analysis of the relationship between spin networks and tensor networks has shown that spin networks can be understood as generalized tensor networks, and in particular as generalized Projected Entangled Pair States (PEPS)~\cite{Colafranceschi:2020ern}. This analysis clarifies the key differences between the two frameworks, namely the presence of gauge symmetry at the nodes, the dynamical nature of the data carried by links and vertices, and the resulting background independence of spin-network states. Related observations had already appeared in earlier works~\cite{Singh_2010,Singh:2017tet,Han:2016xmb}. Recently, Czelusta and Mielczarek~\cite{Czelusta:2025prd} further developed a tensor-network representation of $SU(2)$ spin-network states with applications to holographic bulk--boundary maps. Building on this analogy, holographic properties of spin networks have been explored in several contexts, including the emergence of Ryu–Takayanagi–like entropy formulas and the construction of bulk-to-boundary maps~\cite{Chirco:2017wgl,Chirco:2021chk,Colafranceschi:2021acz,Colafranceschi:2022dig,Chen:2021vrc}. Complementary developments include a quantum-error-correction approach to entanglement entropy in LQG~\cite{Tobin:2025arXiv}, an algebraic derivation of a geometrical area law from entropy maximization for fixed-area states~\cite{Han:2026prd}, and quantum-information characterizations of random spin networks and quantum tetrahedra~\cite{Cepollaro:2023prd,Cepollaro:2023prd2,Cepollaro:2024prd}. In parallel, a number of works have investigated tensor networks with gauge symmetry and their connections to discrete quantum geometries (see e.g.~\cite{Dong:2024jhep,Akers:2024ixq,Balasubramanian:2025rcr}), further strengthening the bridge between these two frameworks.\smallskip

In this work we build on these connections by importing and adapting a tensor-network technique introduced in~\cite{Qi:2017ohu} to the spin-network setting.  The key technical ingredient is an enriched entanglement structure on spin-network links, implemented through additional discrete link variables $\{a_l\}$ that control the amount of entanglement shared between adjacent nodes. The starting point is the observation that, in ordinary spin networks, the gluing of vertices (dual to simplices) is implemented by maximally entangled, gauge-invariant singlet states on the links, while unglued faces correspond to links carrying no entanglement~\cite{Colafranceschi:2020ern}. Working on a complete graph, we promote this binary choice to a controlled, tunable structure by allowing each link to carry a variable amount of entanglement, interpolating between product states and the fully entangled singlet state.

Interestingly, the enrichment of spin-network links with additional entanglement degrees of freedom admits a natural interpretation in terms of edge modes. In gauge theories, edge modes are typically associated with the presence of a boundary: when gauge transformations act non-trivially at the boundary, additional boundary degrees of freedom must be introduced to restore gauge invariance. In tensor-network models based on gauge theories, edge modes can contribute to the area term in boundary entropies~\cite{Donnelly:2016qqt,Dong:2024jhep}. Akers, Soni and Wei~\cite{Akers:2024scipost} construct a model based on topological gauge theory in which edge-mode entanglement is multipartite and the area is not determined by counting cut links. By contrast, in our construction the entanglement carried by each link is bipartite, interpolating between a product state and the singlet. In the spin-network setting, a link is formed by gluing together two semi-links, each dual to a face of a simplex. When the two semi-links are contracted through the maximally entangled, gauge-invariant singlet state, the corresponding faces are fully glued, producing an internal face shared by two simplices and enforcing gauge invariance across the link. By contrast, when the two semi-links are not entangled, they fail to combine into a genuine internal link: the dual faces behave as boundary faces of the respective simplices, and gauge transformations act independently on the two sides. 
Allowing for partially entangled link states interpolates between these two situations. The two semi-links are neither fully glued nor completely independent, and the diagonal gauge invariance associated with an internal face is generically broken. In this sense, intermediate entanglement corresponds to the emergence of boundary-like degrees of freedom localized on the semi-links. These additional, non-gauge-invariant data can be naturally interpreted as edge-mode–like degrees of freedom.

The link-enrichment parameters may also be regarded as local reference frames. The introduction of reference frames in spin-network structures is an active line of research in the context of relational observables in GFT (see~\cite{Calcinari:2024pek,Marchetti:2024nnk} and references therein). Unlike those approaches, where reference frames are typically implemented via scalar or vector fields, used to encode embedding or translational degrees of freedom and attached to nodes, here the additional degrees of freedom live on links and are associated with the local rotational $SU(2)$ symmetry of the spin network. They may thus be interpreted as encoding information about local rotational frames associated with links or semi-links.

Building on this enriched spin-network structure, we focus on the bulk–to-boundary map naturally defined by spin-network contraction, where the input data are the effective link variables $\{a_l\}$, which control how much entanglement is present between the endpoints of each potential link $l$. Different assignments of $\{a_l\}$ interpolate between configurations in which nodes are effectively glued along faces and configurations in which they are effectively disconnected, so that the combinatorial structure of the geometry is not fixed a priori but selected by the entanglement pattern itself.

Given a choice of vertex intertwiners, contracting all internal link degrees of freedom while leaving the boundary legs open defines a linear map from the bulk entanglement data $\{a_l\}$ to a boundary quantum state. In this sense, the bulk–to-boundary map reorganizes information encoded in the pattern of link entanglement into boundary degrees of freedom. The vertex intertwiners do not appear as dynamical degrees of freedom of the output state; rather, they specify how bulk information is routed toward the boundary, acting as fixed tensors that define the map itself. This viewpoint allows us to regard the construction as a family of bulk–to-boundary maps parametrized by the choice of intertwiners, whose typical properties can be probed by averaging over intertwiners with the natural Haar measure.

Within this framework, we find that the bulk–to-boundary map exhibits a controlled holographic behavior. Following the strategy introduced in~\cite{Qi:2017ohu}, we analyze the averaged Rényi entropy of the induced boundary state, and show that in a suitable large-spin regime the map becomes a co-isometry in expectation value, meaning that bulk inner products are preserved on average when mapped to the boundary. Moreover, when attention is restricted to a code subspace defined by configurations in which the entanglement pattern selects a well-defined combinatorial structure, with only controlled fluctuations around it, the map becomes an isometry up to corrections that vanish in the large-spin limit, with its adjoint providing the inverse on the code subspace. 

This leads to a concrete realization of quantum error–correcting features directly within the spin-network Hilbert space. In particular, by analogy with tensor-network models, we expect bulk operators supported on a region selected by the entanglement pattern to be reconstructable from boundary operators acting on an appropriate subset of boundary degrees of freedom, as in entanglement-wedge reconstruction.
Importantly, this realization remains fully discrete and intrinsically geometric: the relevant degrees of freedom retain their interpretation in terms of spin-network data (spins encoding the areas of the dual faces, intertwiners the volume and shape of the dual polyhedra) and link entanglement, rather than being abstract bond dimensions as in the RTN setting.

\paragraph{Organization of the Paper}

In Section~\ref{sec:framework} we introduce the general framework underlying our construction. We begin by reviewing PEPS tensor networks and the idea of enriching the entanglement structure of network links, following techniques originally introduced in the tensor-network literature, together with a discussion of how boundary degrees of freedom are incorporated in this setting. We then adapt these ideas to the spin-network framework. After reviewing the mathematical structure of spin-network vertices and the gluing procedure that enforces gauge invariance, we introduce discrete link entanglement variables and explain how allowing partial entanglement generically breaks gauge invariance at internal links. We conclude the section by defining spin-network boundary nodes and the associated boundary Hilbert space.

Section~\ref{sec:Btb_iso} is devoted to the analysis of the bulk–to-boundary map induced by spin-network contraction. We study its properties using the averaged second Rényi entropy of the induced boundary state. We show that, under suitable conditions on the spins and graph structure, the map becomes a co-isometry in expectation value, and we establish explicit criteria for this behavior in the case of quadrivalent nodes.

In Section~\ref{sec:code_subspace} we introduce a code subspace associated with controlled fluctuations of the link entanglement variables around a fixed entanglement pattern, which defines an effective combinatorial structure of the spin-network graph, with links either effectively present or absent. Within this setting, we analyze when the bulk–to-boundary map becomes an isometry. We then analyze the overlap between boundary states corresponding to distinct bulk configurations and show how it is governed by entropic properties of the region where the configurations differ.

Finally, in Section~\ref{sec:discussion} we discuss the interpretation of our results and their relation to tensor-network models of holography and quantum error correction, highlighting both conceptual similarities and key differences arising from the spin-network setting. We also outline possible extensions and open questions motivated by our construction.

\newpage
\section{General Framework}\label{sec:framework}

\subsection{Tensor networks}
\label{sec:TN}
In this subsection we review the construction of PEPS tensor networks and the enrichment of their entanglement structure, following the presentation of~\cite{Qi:2017ohu}.

\paragraph{Projected Entangled Pair States}
The tensor networks we focus on are projected entangled pair states (PEPS). Their construction proceeds as follows. Consider a graph $\gamma$ with node set $N$ and link set $L$. A link is denoted by an ordered pair $l=(n_i,n_j)$, where an orientation may be chosen for convenience (with $n_i$ the \emph{source} and $n_j$ the \emph{target}).

To each link $l\in L$ we associate a pair of maximally entangled "virtual" qudits of local dimension $D$, described by the Hilbert space $\mathcal{H}_D$. We denote this state by
\[
    \ket{l} \in \mathcal{H}_D \otimes \mathcal{H}_D .
\]
These virtual qudits mediate correlations across the network but do not represent physical degrees of freedom.

Next, for each node $n\in N$ of valence $v_n$, we choose a pure state
\[
    \ket{n} \in \mathcal{H}_n, 
    \qquad 
    \mathcal{H}_n := \bigotimes_{v_n} \mathcal{H}_D ,
\]
which acts as the local tensor contracting the virtual qudits carried by the links incident on $n$. 
Contracting all link pairs with these node tensors produces the PEPS associated with $\gamma$:
\begin{equation}
    \Psi_\gamma 
    = 
    \left( \bigotimes_{n\in N} \bra{n} \right)
      \left( \bigotimes_{l\in L} \ket{l} \right).
    \label{PEPS}
\end{equation}
Since all virtual legs are fully contracted, $\Psi_\gamma$ is a scalar rather than a physical state. Physical degrees of freedom are introduced by attaching \emph{uncontracted} legs to the network, either as additional physical indices at the nodes, or as boundary "dangling" legs connected through an extra link. The resulting tensor network with open legs defines a genuine quantum state whose structure will be discussed at the end of this section.

\begin{figure}
    \centering
    \includegraphics[width=0.7\linewidth]{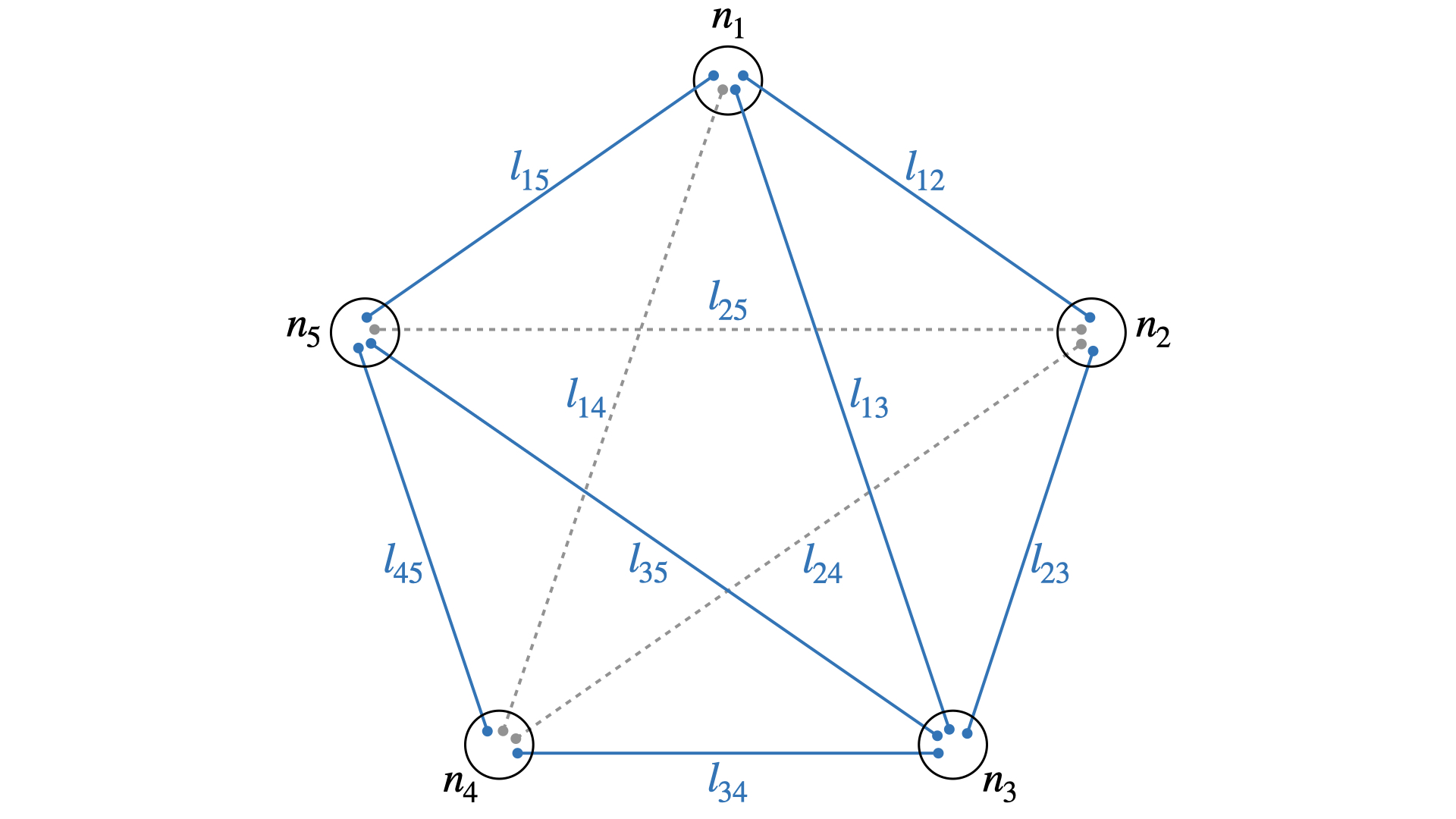}
    \caption{Complete graph on $5$ nodes used to embed all PEPS constructions into a common Hilbert space. Each link carries a pair of qudits, represented by the circular endpoints, while the node tensors that contract these qudits are shown as solid circles. Solid links denote maximally entangled qudit pairs, and dashed links correspond to separable pairs. Node labels $n_i$ and link labels $l_{ij}=(n_i,n_j)$ are shown explicitly.}
    \label{fig:peps}
\end{figure}

Next, we want to define superpositions of tensor networks based on different graphs with the same number of vertices. A difficulty arises from the fact that the dimension of the Hilbert space $\mathcal{H}_n$ associated with a node $n \in N$ depends on its valence. As a result, the definition of the node state $\ket{n}$ would be ambiguous when considering graphs with different valences. To overcome this issue, we construct all node Hilbert spaces as if the underlying graph were the complete graph on $N$ vertices. In other words, we assume each node has $N-1$ virtual neighbors, so that the corresponding Hilbert space is 
$D^{\,N-1}$-dimensional:
\[
    \mathcal{H}_n := \left( \mathcal{H}_D \right)^{\otimes (N-1)} .
\]
The complete-graph embedding is illustrated in Fig.~\ref{fig:peps}. For links that are not actually present in the graph $\gamma$, we replace the maximally entangled pair with a fixed product state of two qudits, denoted by $\ket{l}_0$. With this prescription, every graph $\gamma$ on $N$ nodes gives rise to a well-defined state on the links of the complete graph, independently of the valence structure of $\gamma$. Let $L_N$ denote the set of links of the complete graph on $N$ nodes, with a bijection between its nodes and those of $\gamma$. We can therefore define the link state of $\gamma$ as
\begin{equation}
    \ket{L_\gamma}:=\bigotimes_{l \in L_\gamma} \ket{l}\;\bigotimes_{l \in L_N \setminus L_\gamma} \ket{l}_0 \, .
\end{equation}
Given a graph $\gamma$ and a fixed choice of node states $\{\ket{n}\}_{n \in N}$, the corresponding PEPS can be written in terms of its link state:
\[
    \Psi_\gamma = \left( \bigotimes_{n \in N} \bra{n} \right)
    \ket{L_\gamma} \, .
\]
Thanks to the embedding of all link states into the same Hilbert space, tensor-network states associated with different graphs can be superposed. A superposition of PEPS on two graphs $\gamma$ and $\gamma'$ (with the same node set $N$ and the same node tensors) then takes the form
\begin{equation}
    \Psi = \left( \bigotimes_{n \in N} \bra{n} \right)
    \bigl(
        c_1 \ket{L_\gamma}
        +
        c_2 \ket{L_{\gamma'}}
    \bigr) \, ,
    \label{linear_comb}
\end{equation}
with complex coefficients $c_1$ and $c_2$ (typically normalized as 
$|c_1|^2 + |c_2|^2 = 1$).

\paragraph{Enriching the entanglement structure}

In the construction above, the entanglement structure of the \emph{virtual} degrees of freedom reflects the connectivity of the graph: each link in $L$ carries a maximally entangled pair of $D$-dimensional qudits, contributing $\log D$ units of entanglement, while links not in $L$ contribute none. We now enrich this structure by allowing intermediate configurations between a product state and a maximally entangled state.

\begin{figure}
    \centering
    \includegraphics[width=0.8\linewidth]{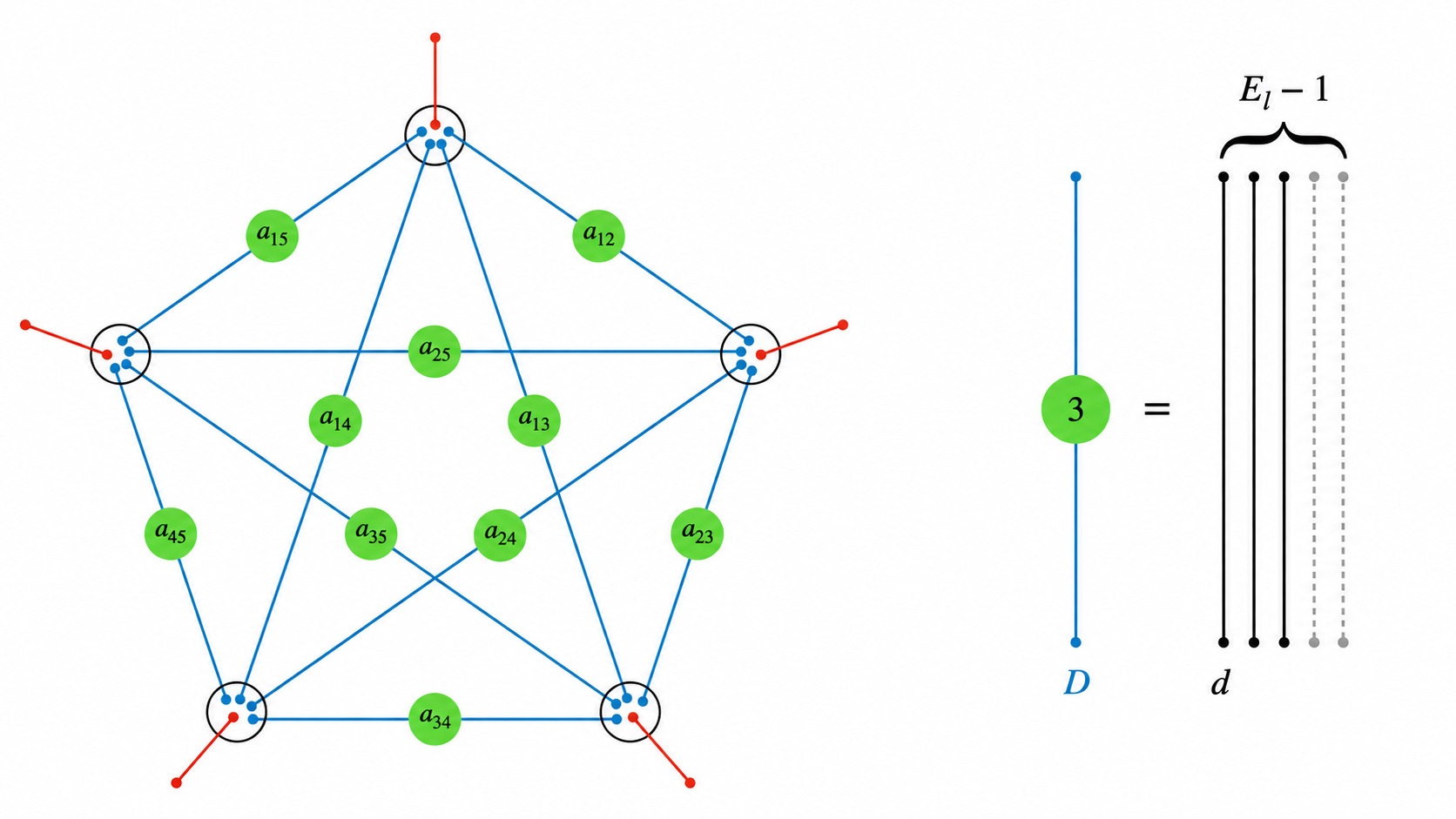}
    \caption{(Left) Complete-graph embedding used to define generalized tensor networks with variable link entanglement. Each link carries an integer $a_l \in \{0,\ldots,E_l-1\}$ that specifies how many of the $(E_l - 1)$ $d$-dimensional qudit pairs are maximally entangled. Boundary nodes are also shown. (Right) Example of a link with $a_l = 3$. Here the original $D$-dimensional qudit is decomposed into $(E_l - 1) = 5$ smaller-$d$ qudits: three of the pairs are maximally entangled (solid), while the remaining two are separable (dashed).}
    \label{fig:peps_a}
\end{figure}

To this end, instead of assigning a single maximally entangled pair on each link, we decompose the $D$-dimensional Hilbert space as
\begin{equation}
\label{eq:H_d}
    \mathcal{H}_D \;\cong\; \bigl(\mathcal{H}_d\bigr)^{\otimes (E_l - 1)},
\end{equation}
where $d\geq 2$ is fixed and $E_l$ is defined by the condition
\begin{equation}
    d^{\,E_l - 1} = D.
\end{equation}
Thus each link $l$ is associated with $(E_l - 1)$ virtual qudit pairs of local dimension $d$. Each pair may be placed either in a product state or in a maximally entangled state. If none of them is entangled, the two virtual endpoints remain unentangled; if all are maximally entangled, the virtual entanglement entropy carried by the link is $(E_l - 1)\log d = \log D$. If $a_l$ out of the $(E_l - 1)$ pairs are maximally entangled, the link carries an entropy $a_l \log d$, interpolating between $0$ and $\log D$. We can then introduce a discrete link variable $a_l \in \{0,1,\ldots,E_l-1\}$ that specifies the number of maximally entangled $d$-dimensional pairs on $l$. An illustration of this decomposition is shown in Fig.~\ref{fig:peps_a} (right). Using Eq.~\eqref{eq:H_d}, the corresponding link state can be constructed as the tensor product of $a_l$ maximally entangled pairs on $\mathcal{H}_d \otimes \mathcal{H}_d$ and $(E_l - 1 - a_l)$ fixed product pairs. In a chosen basis $\{\ket{\alpha}\}$ of $\mathcal{H}_D$, this may be expressed as
\begin{equation}
    \ket{a_l} 
    = 
    \sum_{\alpha,\beta}
    M^{a_l}_{\alpha\beta}\,
    \ket{\alpha}\ket{\beta}
    \;\in\;
    \mathcal{H}_D \otimes \mathcal{H}_D ,
\end{equation}
where the matrix $M^{a_l}$ has Schmidt rank $d^{\,a_l}$, increasing monotonically with the amount of entanglement. The extreme cases can be taken as
\[
    M^{a_l = (E_l - 1)}_{\alpha\beta}
    = \frac{\delta_{\alpha\beta}}{\sqrt{D}},
    \qquad
    M^{a_l = 0}_{\alpha\beta}
    = u_\alpha\, v_\beta,
\]
with $u_\alpha$ and $v_\beta$ unit vectors in~$\mathbb{C}^D$, corresponding respectively to a maximally entangled state and a product state.

A generalized tensor network on the node set $N$ with link entanglement data 
$\{a_l\}_{l \in L_N}$ is then defined as
\begin{equation}
    \Psi(\{a_l\}) 
    = 
    \left( \bigotimes_{n \in N} \bra{n} \right)
    \left( \bigotimes_{l \in L_N} \ket{a_l} \right).
\end{equation}
Standard PEPS are recovered as the special case $E_l = 2$ and $d = D$, for which the only possible values are $a_l = 0$ (product state) and $a_l = 1$ (maximally entangled state).

\paragraph{Boundary nodes}

One may also include a set $B$ of \emph{boundary nodes}. Each boundary node $b \in B$ is connected to exactly one bulk node 
$n_b \in N$ by a maximally entangled pair of qudits (see Fig.~\ref{fig:peps_a}, left), represented by the state $\ket{l_b} \in \mathcal{H}_b \otimes \mathcal{H}_b$. Here $\mathcal{H}_b$ denotes the one-qudit Hilbert space associated with the boundary degree of freedom; its dimension is arbitrary and may differ from the bulk qudit dimension $D$, which is why we adopt a distinct notation.  
A boundary link increases the valence of $n_b$, so we extend its Hilbert space by tensoring with the additional boundary space $\mathcal{H}_b$, such that
\[
   \ket{n_b} \in \mathcal{H}_{n_b} \otimes \mathcal{H}_b .
\]
Including the boundary nodes, the full tensor-network takes the form
\begin{equation}
    \ket{\Psi} =
    \left( \bigotimes_{n \in N} \bra{n} \right)
    \left( \bigotimes_{l \in L_N} \ket{a_l} \right)
    \left( \bigotimes_{b \in B} \ket{l_b} \right)
\end{equation}
Note that, after contracting all bulk degrees of freedom, the resulting state lives in the boundary Hilbert space $\mathcal{H}_B :=\bigotimes_{b \in B} \mathcal{H}_b$.

\subsection{Spin networks}
\label{sec:SN}
\paragraph{The spin network vertex} A spin network is a graph whose links are labelled by irreducible unitary representations of a Lie group $G$, and whose vertices are labelled by $G$-invariant tensors (intertwiners). When the graph is taken to be dual to a simplicial decomposition of a spatial slice, each vertex represents an elementary ``chunk'' of space (a simplex), while each link corresponds to a face 
shared by two neighboring chunks. The representation labels encode the areas of the dual faces, and the intertwiners encode the volume and shape degrees of freedom of the dual simplices. In what follows, we specialize to the case $G = SU(2)$ and restrict attention to four-valent vertices, which naturally correspond to tetrahedra in the dual picture. The construction, however, generalizes straightforwardly to $k$-valent nodes; in fact, the discrete geometric interpretation of spin networks extends beyond the simplicial case to generic 3d polyhedra~\cite{Bianchi:2010gc}, and so does the gluing by entanglement.

\begin{figure}
    \centering
    \includegraphics[width=0.8\linewidth]{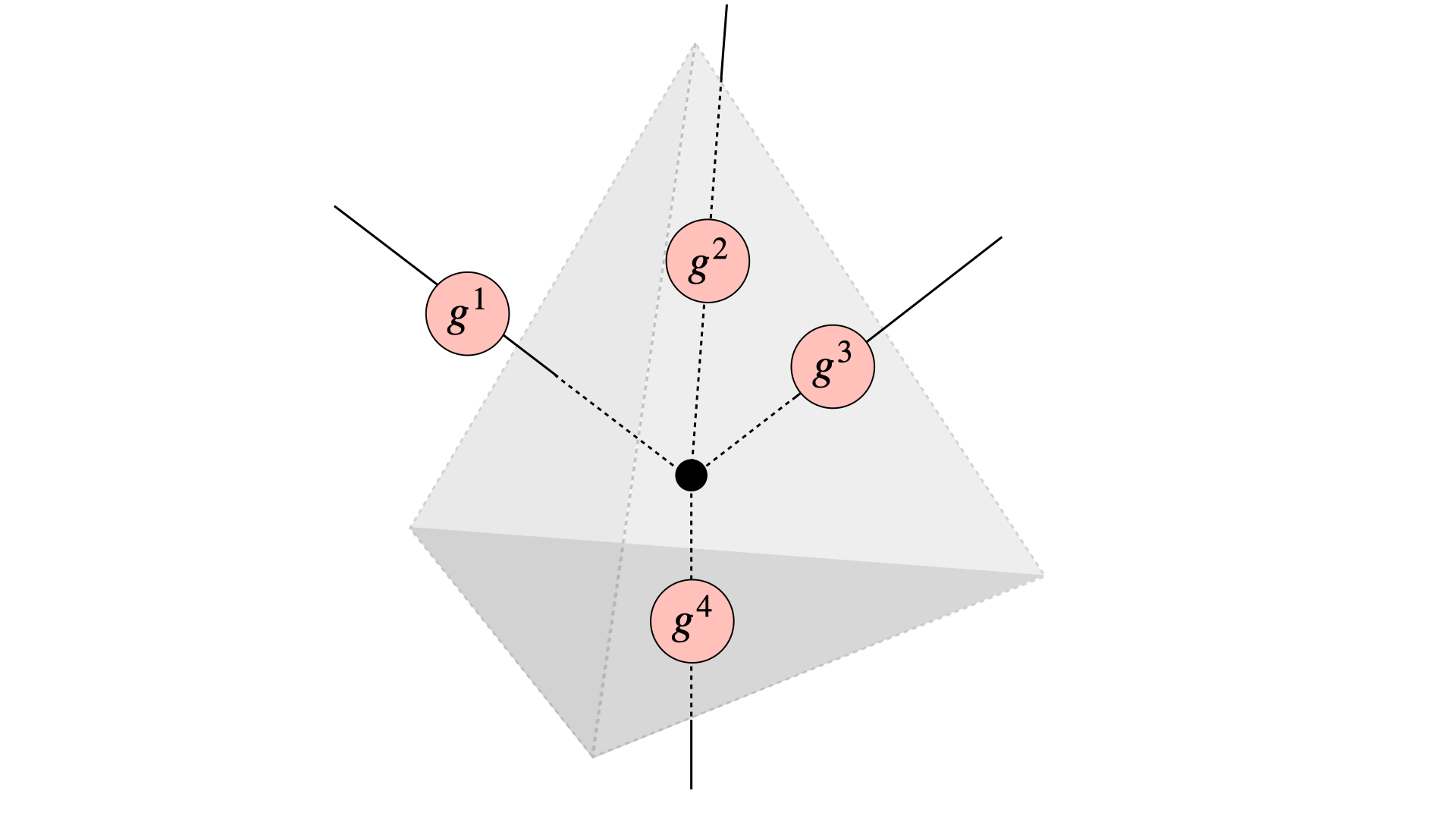}
    \caption{A four-valent spin-network vertex. Each open link carries an $SU(2)$ group element $g^i$, and the center solid node represents the intertwiner. The spin-network vertex is dual to a tetrahedron embedded in three-dimensional space, with link group elements dual to its four surfaces and the intertwiner dual to its interior.}
    \label{fig:vertex_g}
\end{figure}

Let us begin by analyzing the mathematical structure of a single spin network vertex, before explaining how many such vertices assemble into a graph. A four-valent vertex is depicted in Fig.~\ref{fig:vertex_g}. Each of its incident links carries a group element $g_i \in SU(2)$, interpreted as the holonomy of the Ashtekar connection along that link. The Hilbert space associated with the vertex, describing the quantum geometry of the dual tetrahedron, is\footnote{This structure is best understood by recalling the corresponding classical picture. Consider a tetrahedron in Euclidean three-dimensional space, whose faces are labelled by $i=1,\dots,4$. Its classical geometry can be described by four vectors $\{\vec{L}_i\}_{i=1}^4 \in \mathbb{R}^3 \simeq \mathfrak{su}(2)$, where each $\vec{L}_i$ is normal to the $i$-th face and has norm equal to the face area. These vectors satisfy the \emph{closure constraint} $\sum_{i=1}^4 \vec{L}_i = 0$, ensuring that the faces close to form a polyhedron (the geometry is defined up to global rotations). Since the $\vec{L}_i$ belong to the Lie algebra $\mathfrak{su}(2)$, one may equivalently encode the same geometry using group elements $\{g^i\}_{i=1}^4 \in SU(2)$, up to a common left action of $SU(2)$ which implements the closure constraint. Quantising this classical phase space leads precisely to the Hilbert space $L^2(SU(2)^4/SU(2))$.}
\begin{equation}
   \mathcal{H} = L^2\!\big(SU(2)^4/SU(2)\big),
\end{equation}
whose elements are square-integrable wave functions $f(\vec{g}) = f(g^1,g^2,g^3,g^4)$ satisfying the gauge-invariance condition
\begin{equation}
    f(\vec{g}) = f(h\vec{g}) \qquad \forall\, h \in SU(2),
    \label{invariance}
\end{equation}
with $h\vec{g} := (h g^1, h g^2, h g^3, h g^4)$. That is, the wavefunction is invariant under a global rotation $h$ acting simultaneously on all four links; the quotient by $SU(2)$ in $L^2(SU(2)^4/SU(2))$ reflects precisely this redundancy.

\begin{figure}
    \centering
    \includegraphics[width=0.8\linewidth]{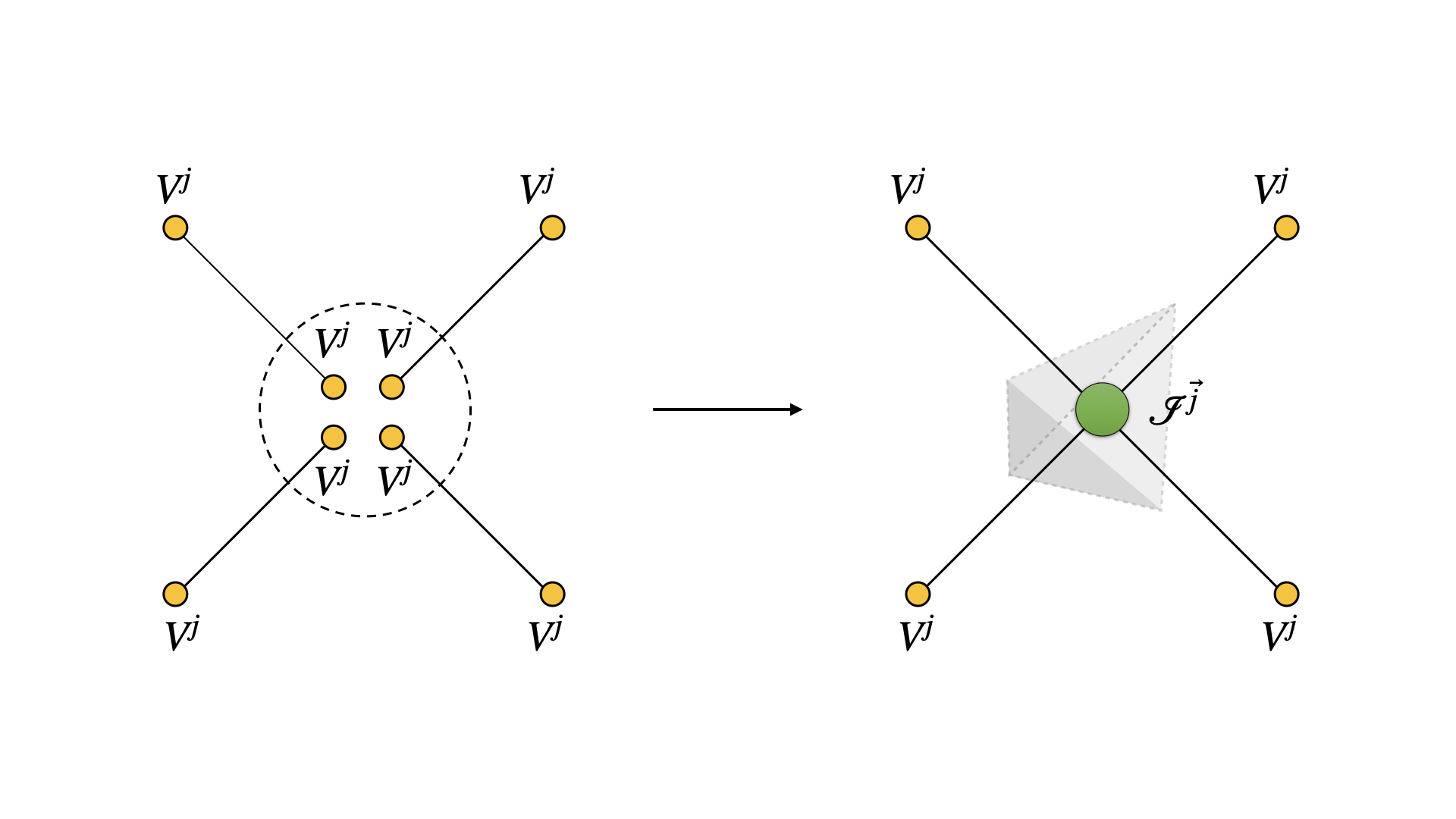}
    \caption{\textbf{Left:} Each link carries a representation space $V^{j_i}$ at each of its endpoints, depicted as yellow circles and corresponding to the representation label $j_i$ in the Peter–Weyl decomposition. The four representation spaces meeting at the vertex are collectively projected onto the $SU(2)$-invariant subspace, indicated by the dashed black circle and corresponding to the recoupling map in Eq.~\eqref{eq:recoupling}. \textbf{Right:} After imposing gauge invariance, the four representation spaces are recoupled into a single intertwiner $\mathscr{I}^{\vec j}$ (green circle), representing an $SU(2)$-invariant tensor associated with the vertex. The external yellow circles denote the remaining representation spaces carried by the links.}
    \label{fig:vertex_j}
\end{figure}

By the Peter-Weyl theorem, the wave functions admit an expansion in terms of matrix elements of irreducible representations of $SU(2)$:
\begin{equation}
    f(\vec{g})
    =
    \sum_{\vec{j},\,\vec{m},\,\vec{n}}
        f^{\vec{j}}_{\vec{m}\vec{n}}
        \prod_{i=1}^{4} \sqrt{2j^i + 1}
        D^{j^i}_{m^i n^i}(g^i),
        \label{pw}
\end{equation}
where $j^i \in \tfrac{\mathbb{N}}{2}$ and $m^i,n^i$ label bases of the 
representation space $V^{j^i}$ and its dual, and $D^{j^i}_{m^i n^i}(g^i)$ are the corresponding Wigner matrix elements. At the level of Hilbert spaces, this corresponds to the decomposition (illustrated in Fig.~\ref{fig:vertex_j})
\begin{equation}
    L^2(SU(2)^4)
    \;=\;
    \bigotimes_{i=1}^4 
        \widehat{\bigoplus_{j^i=0}}
            \left(
                V^{j^i} \otimes V^{j^i}
            \right),
\end{equation}
where the hat denotes completion with respect to the $L^2$ norm; in the 
following we omit the hat for notational simplicity. The quotient by $SU(2)$, 
i.e.\ the implementation of the gauge invariance condition~\eqref{invariance}, 
has not yet been made explicit. As can be seen by decomposing~\eqref{invariance} using 
the Peter-Weyl theorem, gauge invariance amounts to projecting the 
representation spaces $V_{j^i}$ onto the subspace of $SU(2)$-invariant tensors (see Fig.~\ref{fig:vertex_j}):
\begin{equation}
    \label{eq:recoupling}
    V^{j^1}\otimes \cdots \otimes V^{j^4}
\;\longrightarrow\;
\mathrm{Inv}_{SU(2)}
        \!\left[V^{j^1}\otimes \cdots \otimes V^{j^4}\right]
        \;=:\; \mathscr{I}^{\vec{j}}.
\end{equation}
The vertex Hilbert space therefore admits the decomposition
\begin{equation}
    \mathcal{H}
    = L^2\!\big(SU(2)^4/SU(2)\big)
    = \bigoplus_{\vec{j}}
        \left(
            \mathscr{I}^{\vec{j}}
            \otimes
            \bigotimes_{i=1}^4 V^{j^i}
        \right),
    \label{H_vertex}
\end{equation}
and the corresponding wave function takes the form
\[
    f(\vec{g})
    =
    \sum_{\vec{j},\,\vec{n},\,\iota}
        f^{\vec{j}}_{\vec{n}\iota}\,
        \chi^{\vec{j}}_{\vec{n}\iota}(\vec{g}),
\]
where $\iota$ labels a basis of the intertwiner space $\mathscr{I}^{\vec{j}}$, 
and $\chi^{\vec{j}}_{\vec{n}\iota}$ are the spin-network basis functions for a 
single four-valent vertex:
\begin{equation}
\chi^{\vec{j}}_{\vec{n},\iota}(\vec{g})
\;\coloneqq\;
\iota_{\vec{m}}
\prod_{i=1}^4
    \sqrt{2j^i + 1}\,
    D^{j^i}_{m^i n^i}(g^i),
\label{spinwf}
\end{equation}
with $\iota_{\vec{m}}\in \mathscr{I}^{\vec{j}}$ an invariant tensor.

\paragraph{Gluing vertices}
We now proceed to construct a full spin network by gluing initially disjoint vertices along pairs of open links. What does it mean to glue two vertices along a pair of such links? Recall that the group element $g_l$ associated with an oriented link $l$ is the holonomy of the Ashtekar connection along that link. Thus, when gluing a link $l$ of one vertex to a link $l'$ of another vertex, the resulting link of the combined graph must carry the composed holonomy $g_l g_{l'}^{-1}$, where the inverse accounts for the opposite orientation of the second link. Equivalently, a simultaneous left multiplication by \(h \in SU(2)\) at the two open ends leaves this combination invariant: $h g_l\, (h g_{l'})^{-1} = g_l g_{l'}^{-1}.$ This observation shows that the gluing operation can be implemented by acting with the same group element $h$ on the two open legs and then averaging over $h$. To illustrate this explicitly, consider two four-valent vertices with wavefunctions $\Psi_1\in \mathcal{H}_1$ and $\Psi_2\in \mathcal{H}_2$ (where $\mathcal{H}_i$ is the vertex Hilbert space of Eq.~\eqref{H_vertex}). We glue their fourth open legs by applying the same $h$ on those legs and integrating over it. As shown in Appendix~\ref{app:gluing}, this produces the new wavefunction
\begin{equation}
    \int dh \Psi_1(g^1,\dots,g^4h) \Psi_2(q^1,\dots,q^4h) =: \Psi(g_1,g_2,g_3,g,q_1,q_2,q_3)
\end{equation}
where $g=g_4 q_4^{-1}$ is the holonomy assigned to the new internal link. The resulting wavefunction thus depends on seven holonomies: one for each link of the glued two-vertex graph.

\begin{figure}
    \centering
    \includegraphics[width=0.8\linewidth]{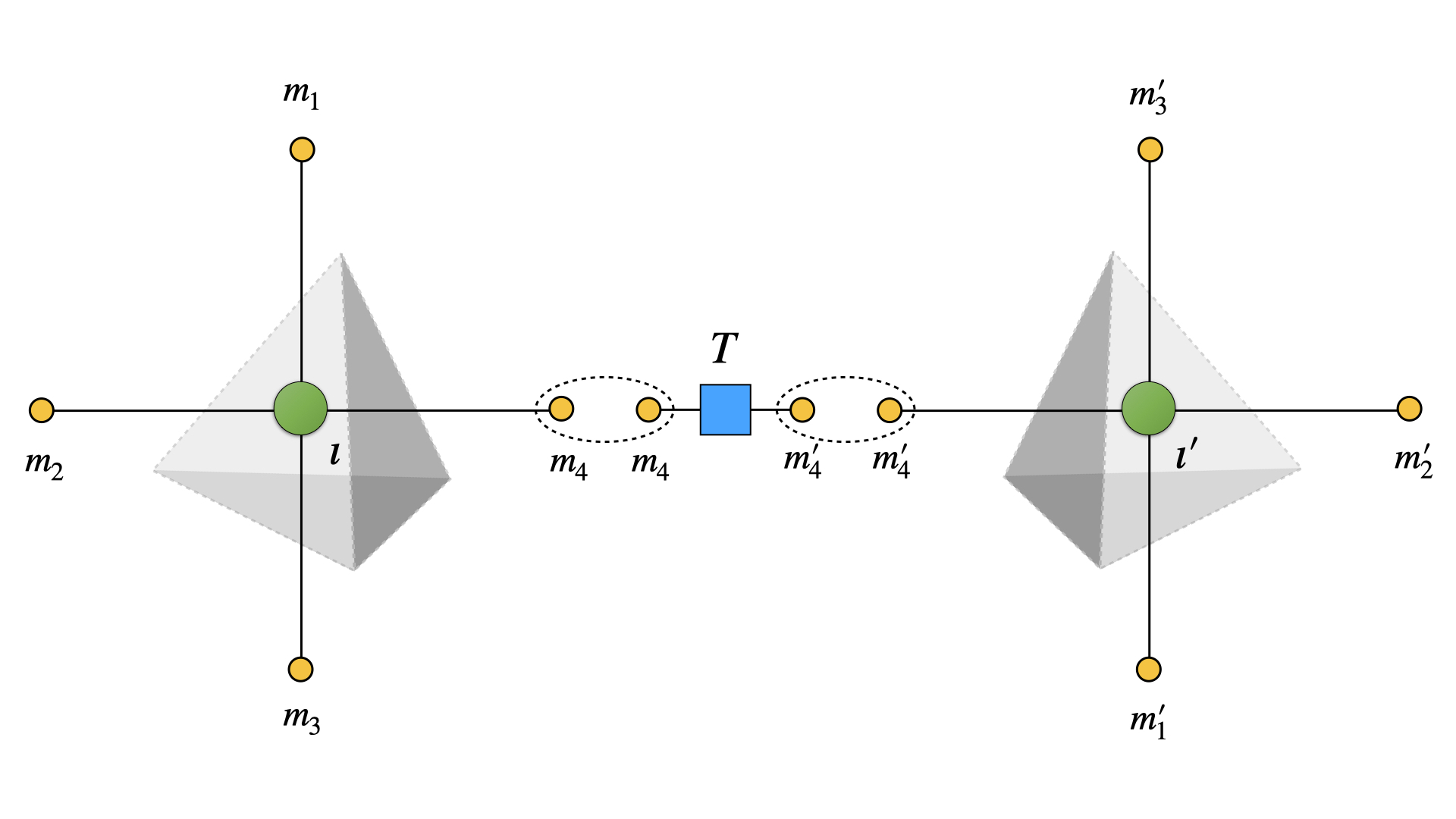}
    \caption{The process of gluing together a pair of open links from two spin-network vertices. From the representation basis perspective, since the resulting link has to be gauge-invariant, this process is attaching a two-valent intertwiner $T$ whose two legs have magnetic numbers $m_4$ and $m'_4$. The only such intertwiner is proportional to the Kronecker delta $\delta_{m_4,-m'_4}.$}
    \label{fig:gluing}
\end{figure}

In the representation basis, the gluing operation corresponds to the contraction of the two open legs with the unique bivalent intertwiner. Indeed, in Appendix~\ref{app:gluing} we show that
\begin{multline}
\int dh \Psi_1(g^1,\dots,g^4h) \Psi_2(q^1,\dots,q^4h)\\= \left({\Psi_1}^{\vec{j}}_{\vec{m}\iota}{\Psi_2}^{\vec{j}'}_{\vec{m}'\iota'}T_{m_4m_4'}\right)\left(\chi^{\vec{j}}_{m_1m_2m_3k,\iota}(\vec{g}) \chi^{\vec{j}}_{m_1'm_2'm_3'k',\iota'}(\vec{q})T_{kk'}\right)
\label{gluing}
\end{multline}
where $T_{kk'}\in V^{j}\otimes V^{j}$ is the $SU(2)$-invariant tensor
\begin{equation}\label{bivalent}
T_{kk'}\coloneqq \frac{(-1)^{j+k}}{\sqrt{2j+1}}\delta_{k,-k'}
\end{equation}
Thus, gluing two open legs is implemented by contracting the corresponding magnetic indices with the invariant tensor $T_{k k'}$. In physical terms, the operation enforces the $SU(2)$ gauge invariance at the newly formed internal link by projecting onto the unique spin-0 (singlet) intertwiner. This is illustrated in Fig.~\ref{fig:gluing}. \medskip 

The gluing procedure described above makes explicit that identifying two open spin-network legs amounts to imposing maximal entanglement (within each $j$-sector) between the corresponding degrees of freedom. This is precisely the mechanism underlying the construction of PEPS. The difference is merely one of viewpoint: in PEPS one typically starts from entangled link states and contracts them with local tensors, whereas our spin-network construction proceeds by first specifying the vertex tensors (intertwiners) and then gluing them.

Indeed, the gluing of two open legs of representation $j$ can be implemented by contracting them with the unique $SU(2)$-invariant state (the singlet state) in $V^j\otimes V^j$:
\begin{equation}
    \ket{l_j} := \frac{1}{\sqrt{2j+1}}\sum_{m=-j}^{m=j} (-1)^{j+m}\ket{jm}\ket{j,-m}
\end{equation}
To be more general, we may introduce the link state
 \begin{equation}
    \ket{l} := \bigoplus_j  \frac{1}{\sqrt{2j+1}}\sum_{m=-j}^{m=j} (-1)^{j+m}\ket{jm}\ket{j,-m} ~\in ~\mathcal{H}_l := \bigoplus_j V^j\otimes V^j
    \label{link-SN}
\end{equation}
which simultaneously includes all j-sectors and plays the role of the virtual maximally entangled pair familiar from PEPS.
Thus, if $\{\ket{n}\}_{n\in N}$ denote the spin-network vertex states and $\{\ket{l}\}_{l\in L}$ the link states, a spin network on a graph with nodes $N$ and links $L$ can be written compactly as
\begin{equation}
    \ket{\Psi} = \left(\bigotimes_{l \in L}\bra{l}\right) \left( \bigotimes_{n \in N} \ket{n}\right) 
\end{equation}
This expression mirrors the PEPS formula of Eq.~\eqref{PEPS}: each link is endowed with a maximally entangled state, and each vertex with a gauge-invariant tensor enforcing the closure constraint.

\paragraph{Link entanglement variables and the breaking of gauge invariance}

The link state~\eqref{link-SN} in the spin-network formalism already allows for superpositions of different graphs, thanks to the presence of the $j=0$ sector. Indeed, when $j=0$ the sum over magnetic indices contains only a single term,
corresponding to a one-dimensional Hilbert space and therefore to a \emph{non-entangled} link. Geometrically, $j=0$ implies that the two tetrahedra represented by the adjacent vertices share a face of zero area, meaning that they are not adjacent in the sense of Regge geometry.  

This implies that when comparing or superposing spin networks based on different graphs, one may embed all of them into a common larger graph (e.g.~the complete graph on $N$ vertices) and assign $j=0$ to the missing links. This is very close to the procedure based on cylindrical equivalence, used in canonical LQG to define the full kinematical Hilbert space of the theory and the Ashtekar--Lewandowski measure~\cite{Ashtekar:2004eh}. The resulting formalism is general enough to handle arbitrary node valence, just as in tensor networks; when one restricts to 4-valent vertices, these maintain the familiar interpretation as duals of tetrahedra.

Working on the complete graph, the Hilbert space at a vertex becomes

\begin{equation}
    \mathcal{H}_n
    = \bigoplus_{j_1 \dots j_{N-1}}
        \left(
            \mathscr{I}^{j_1 \dots j_{N-1}}
            \otimes
            \bigotimes_{i=1}^{N-1} V^{j^i}
        \right),
    \label{H_vertex_complete}
\end{equation}
Then one can impose the condition that only four legs carry nonzero spin in order to recover the geometric interpretation of a tetrahedron.

Following the tensor-network construction, we now enrich the space of link states by allowing intermediate entanglement between the two endpoints of each link.

Suppose a link carries spin $j$. The Hilbert space on each endpoint is the $(2j+1)$-dimensional representation space $V^j$, so the effective dimension of the corresponding PEPS-like qudit is $D_j := 2j + 1$. As in the tensor-network case, we subdivide this qudit into $(E_l(j)-1)$ elementary $d$-dimensional 
qudits, choosing $E_l(j)$ such that
\[
    d^{\,E_l(j)-1} = D_j.
\]
We allow different links to carry different spins, and therefore different dimensions $D_j$, while keeping the elementary dimension $d$ fixed. We then introduce a discrete link variable 
\[
    a_{j_l} \in \{0,1,\dots,E_l(j)-1\},
\]
which specifies the number of elementary $d$-dimensional pairs that are in a maximally entangled state along link $l$. In this way, the entanglement entropy between the two endpoints of the link ranges from $0$ (no entangled pairs) to $\log(2j+1)$ (all pairs maximally entangled), in steps of $\log d$.

Concretely, once we consider the isomorphism $V^j \cong \mathcal{H}_d^{\otimes (E_l(j)-1)}$, we construct the link state $|a_{j_l}\rangle$ as the tensor product of $a_{j_l}$ maximally entangled pairs on $\mathcal{H}_d\otimes\mathcal{H}_d$ and $(E_l(j)-1-a_{j_l})$ fixed product pairs. In a basis $\{\ket{jm}\}$ of $V^j$, this can be written as
\begin{equation}
    |a_{j_l}\rangle
    =
    \sum_{m,n=-j}^{j}
    T^{a_{j_l}}_{mn}\,
    |jm\rangle\otimes|jn\rangle
    \;\in\;
    V^j\otimes V^j,
\end{equation}
where the matrix $T^{a_{j_l}}$ has Schmidt rank $d^{\,a_{j_l}}$, increasing monotonically with the amount of entanglement. The extreme cases can be chosen as
\[
    T^{(a_{j_l}=E_l(j)-1)}_{mn} = T_{mn},
    \qquad
    T^{(a_{j_l}=0)}_{mn} = u_m\,v_n,
\]
where $T_{mn}$ is the bivalent intertwiner defined in Eq.~\eqref{bivalent}, and $u_m,v_n$ are unit vectors in $V^j$. Thus $a_{j_l}=E_l(j)-1$ reproduces the maximally entangled singlet state $|l_j\rangle$, while $a_{j_l}=0$ corresponds to a product state with no entanglement.

A spin-network state on a graph with link spins $\{j_l\}$ and entanglement data $\{a_{j_l}\}$ is then constructed as
\begin{equation}
    |\Psi(\{a_{j_l}\}) \rangle
    =
    \left( \bigotimes_{l\in L} \bra{a_{j_l}} \right)
    \left( \bigotimes_{n\in N} |n\rangle \right),
\end{equation}
in direct analogy with the generalized PEPS construction.\smallskip

It is important to stress that the original link state $|lj\rangle$, defined in Eq.~\eqref{link-SN}, is the unique state in $V^j\otimes V^j$ that is both maximally entangled and invariant under the diagonal action of $SU(2)$:
\[
    (U\otimes U)\,|l_j\rangle = |l_j\rangle
    \qquad \forall\, U\in SU(2).
\]
Equivalently, $|l_j\rangle$ lives entirely in the total-spin $J=0$ subspace of $V^j\otimes V^j \cong \bigoplus_{J=0}^{2j} V^J$, and therefore implements a bivalent intertwiner. By contrast, a generic entangled state $|a_{j_l}\rangle$ constructed as above has nontrivial support on total-spin sectors $J>0$, and is therefore \emph{not} invariant under a diagonal group action. In this sense, replacing the singlet $|l_j\rangle$ with a generic $|a_{j_l}\rangle$ breaks the gauge invariance associated with the diagonal $SU(2)$ at that link: the internal link now carries a nontrivial total spin, rather than implementing a pure intertwiner. Only the special value $a_{j_l}=E_l(j)-1$, for which $|a_{j_l}\rangle = |l_j\rangle$, preserves the gauge invariance at the link.

We now consider the more general situation in which a link may carry a superposition of different spin representations. For each spin $j$, let $|\phi_j\rangle$ denote an arbitrary maximally entangled state in $V^{j}\otimes V^{j}$. All such states have the same entanglement entropy $\log(2j+1)$, although they differ by local unitary transformations acting on a single endpoint. A general link state can therefore be written as a superposition over spin sectors,
\begin{equation}
    |l\rangle = \bigoplus_{j_i} c_i\, |\phi_{j_i}\rangle,
    \label{linear_superposition}
\end{equation}
with $\sum_i |c_i|^2 = 1$. Since the states $|\phi_{j_i}\rangle$ live in mutually orthogonal spin sectors, the reduced density matrix of one endpoint is block-diagonal, with eigenvalues
\[
    \lambda_{j_i} = \frac{|c_i|^2}{2j_i+1}
\]
of degeneracy $(2j_i+1)$. The entanglement entropy of the link is therefore
\begin{equation}
    S_l
    =
    \sum_i |c_i|^2 \log(2j_i+1)
    \;-\;
    \sum_i |c_i|^2 \log |c_i|^2,
    \label{eqn:superposed_entropy}
\end{equation}
as derived in Appendix~\ref{app:superposed}. We then define an effective integer $E_l$ through
\begin{equation}
    (E_l - 1)\,\log d = S_l ,
    \label{dimension}
\end{equation}
so that a link consisting of $(E_l - 1)$ elementary $d$-dimensional qudits in a fully entangled configuration carries entropy $S_l$. Next, we introduce a discrete variable $a_l \in \{0,1,\dots,E_l-1\}$, which specifies how many of these elementary qudits are taken to be maximally entangled. The resulting partially entangled link state $|a_l\rangle$ lives in
\begin{equation}
\mathcal{H}_l=\bigoplus_j V^{j}\otimes V^{j}    
\end{equation}
and has entropy $a_l \log d$, interpolating between a product state ($a_l = 0$) and the fully entangled state ($a_l = E_l - 1$).

At this point it is important to note that gauge invariance at the link is \emph{not} guaranteed for a generic maximally entangled representative $|\phi_j\rangle$. Indeed, such states typically have support in total-spin sectors $J>0$. Gauge invariance is recovered in the fully entangled case $a_l = E_l - 1$ \emph{only if}, for each $j$, the chosen representative $|\phi_j\rangle$ is the $SU(2)$-invariant singlet. In this way, the parameter $a_l$ provides a unified and discrete control of the entanglement structure that remains compatible with superpositions across spin sectors, while leaving the choice of whether (or not) to preserve gauge invariance to the selection of the representatives $|\phi_j\rangle$.

Although the link Hilbert space is in principle given by $\mathcal H_l=\bigoplus_j V^j\otimes V^j$, in the following we do not treat the spin labels as dynamical variables. Superpositions over spin sectors are only used to motivate the definition of an effective entanglement dimension $E_l$. Once $E_l$ is fixed, we restrict attention to the subspace spanned by the partially entangled states $|a_l\rangle$, which provides an effective description of the link degrees of freedom relevant for the bulk-to-boundary map.

\paragraph{Spin network boundary}

Analogously to the tensor-network setup, we introduce a set $B$ of boundary nodes, each having a single open leg. Every boundary node $b\in B$ is connected to exactly one bulk node $n_b\in N$ and carries the Hilbert space $\mathcal{H}_b = \bigoplus_j V^j$.  

A crucial difference with respect to PEPS now appears. In PEPS, the Hilbert space of a vertex factorises over its legs, so attaching a new boundary qudit simply amounts to tensoring an additional leg Hilbert space. Spin-network vertices, however, do \emph{not} factorise over legs: after imposing $SU(2)$ gauge invariance, a $k$-valent vertex has Hilbert space
\[
    \mathcal{H}_{n}
    =
    \bigoplus_{j_1,\dots,j_{k}}
        \left(
            \mathscr{I}^{j_1 \dots j_{k}}
            \otimes
            \bigotimes_{i=1}^{k} V^{j_i}
        \right),
\]
which factorises only within each fixed spin sector, and not across the full direct sum over $\{j_i\}$. The intertwiner space $\mathscr{I}^{j_1\dots j_k}$ couples all legs simultaneously, preventing any PEPS-like decomposition.

For this reason, attaching a boundary node does \emph{not} produce a tensor product $\mathcal{H}_{n_b}^{\text{(old)}} \otimes \mathcal{H}_b$. Instead, increasing the valence of $n_b$ modifies the intertwiner structure, and the correct Hilbert space of the bulk node after adding the boundary leg is
\begin{equation}
    \mathcal{H}_{n_b}
    =
    \bigoplus_{j_1\dots j_{N}}
        \left(
            \mathscr{I}^{j_1 \dots j_{N}}
            \otimes
            \bigotimes_{i=1}^{N} V^{j_i}
        \right),
        \label{eq:H_b}
\end{equation}
where $N$ is the new valence (including the boundary link). Thus the newly added leg participates in the same gauge-invariant coupling as all the other legs.

The link connecting $b$ and $n_b$ carries a general superposition of maximally entangled states of the form introduced previously, and the resulting state is obtained by contracting all internal legs of the bulk graph:
\begin{equation}
    \ket{\Psi(\{a_l\})}
    =
    \left(
        \bigotimes_{l \in L}\bra{a_l}
    \right)
    \left(
        \bigotimes_{n \in N}\ket{n}
    \right)
    \left(
        \bigotimes_{b \in B}\ket{l_b}
    \right),
    \label{bdy_state}
\end{equation}
Note that the resulting state lives in the Hilbert space 
\begin{equation}
    \left(\bigotimes_{n \in N} \mathscr{I}_n \right) \otimes \left(\bigotimes_{b\in B} \mathcal{H}_b\right)
\end{equation}
where 
\begin{equation}
 \mathscr{I}_n = \bigoplus_{\vec{j}}  \mathscr{I}^{\vec{j}} \qquad \mathcal{H}_b=\bigoplus_{j_b} V^{j_b}
\end{equation}

\section{Boundary-to-Bulk Isometry}\label{sec:Btb_iso}

In this section, we view the spin-network construction as defining a bulk-to-boundary map that assigns to each choice of bulk link variables $\{a_l\}$ a boundary state $\ket{\Psi(\{a_l\})}$. The vertex intertwiners do not appear as degrees of freedom of the output state, but rather play the role of parameters specifying the map itself.

The bulk degrees of freedom are encoded in the effective bulk Hilbert space
\begin{equation}
\mathcal{H}_{\mathrm{bulk}}
:= \mathrm{span}\!\left\{
\bigotimes_{l\in L} |a_l\rangle
\right\}
\;\subset\;
\mathcal{H}_L
:= \bigotimes_{l\in L} \mathcal{H}_l,
\qquad
\mathcal{H}_l := \bigoplus_{j_l} V^{j_l}\otimes V^{j_l},
\end{equation}
where $|a_l\rangle$ denotes a partially entangled link state with fixed effective entanglement dimension $E_l$. The boundary Hilbert space is
\begin{equation}
   \mathcal{H}_B = \bigotimes_{b\in B} \mathcal{H}_b,
\qquad
\mathcal{H}_b = \bigoplus_{j_b} V^{j_b}. 
\end{equation}
We also fix a reference boundary state
$\ket{l_\partial}:=\bigotimes_{b\in B}\ket{l_b}\in\mathcal{H}_B$.

For a given choice of vertex intertwiners $\{\,|n\rangle\,\}_{n\in N}$, the spin-network contraction defines a linear map
\begin{equation}
V_{\{n\}}:\ \mathcal{H}_{\mathrm{bulk}} \longrightarrow \mathcal{H}_B,
\qquad
V_{\{n\}}\ket{\{a_l\}} = \ket{\Psi(\{a_l\})},
\label{eq:V_def}
\end{equation}
with
\begin{equation}
\ket{\Psi(\{a_l\})}
=
\left(
\bigotimes_{l \in L}\bra{a_l}
\right)
\left(
\bigotimes_{n \in N}\ket{n}
\right)
\left(
\bigotimes_{b \in B}\ket{l_b}
\right).
\label{Psi(a_l)}
\end{equation}
In this formulation, the intertwiners $\{ |n\rangle \}$ parametrize the bulk-to-boundary map, while the output state lives entirely in the boundary Hilbert space. In the following sections, the isometry property of $V_{\{n\}}$ is established in expectation value by averaging over the intertwiners, in close analogy with random tensor network constructions. \medskip

By definition, $V^\dagger$ is an isometry if
\begin{equation}
    VV^\dagger = \mathbbm{1}_{\mathcal H_{B}}
\end{equation}
and we could rewrite $VV^\dagger$ by using the resolution of the identity on $\mathcal{H}_{\mathrm{bulk}}$ in the $\{\ket{\{a_l\}}\}$ basis:
\begin{equation}
    VV^\dagger=\sum_{\{a_l\} } V\ket{\{a_l\}}\bra{\{a_l\}}V^\dagger=\sum_{\{a_l\} } \ket{\Psi(\{a_l\})}\bra{\Psi(\{a_l\})}
\end{equation}
so the isometry condition becomes
\begin{equation}
    \sum_{\{a_l\} } \ket{\Psi(\{a_l\})}\bra{\Psi(\{a_l\})} = \mathbbm{1}_{\mathcal H_B}
    \label{isometry'}
\end{equation}
The left-hand side of Eq.~\eqref{isometry'} can be viewed as the density matrix $\rho_B$ of some mixed state in the boundary Hilbert space. In addition, the notion of isometry can be generalized to include maps that don't exactly preserve inner product but instead always scale them by a fixed constant $C \in \mathbb{C}$, which allows us to work with non-normalized states. Taking $C=1$ recovers the original definition. Thus what we want to show is
\begin{equation}
    \rho_B \coloneqq \sum_{\{a_l\} } \ket{\Psi(\{a_l\})}\bra{\Psi(\{a_l\})} = C\mathbbm{1}_{\mathcal H_B}
    \label{isometry}
\end{equation}
This equation is equivalent to the statement that $\rho_B$ is maximally mixed, which we can check by computing its entropy. 

\subsection{Second Rényi Entropy}
Instead of the map resulting from a single choice of intertwiners, we compute the entropy averaged over all intertwiners. Since each intertwiner space $\mathscr{I}_n$ is compact, we use the uniform probability measure $\mu_n$ to define the local average $\avgn{-}$, and take the total average $\avg{-}$ as the product over all nodes. We also fix the boundary links to carry a homogeneous spin $j$, so that the boundary node Hilbert space $\mathcal{H}_b$ has dimension $D = 2j+1$ and does not involve a superposition over spin representations.

We characterize the maximal mixing of $\rho_B$ using the second Rényi entropy $S_2(\rho_B)$. The use of the second Rényi entropy is standard in random tensor network constructions, where it provides a technically convenient probe of maximal mixing via the replica (swap) trick. Moreover, as shown in Sec.~3.2 and Appendix~A of~\cite{Qi:2017ohu}, maximality of the second Rényi entropy, together with mild additional assumptions, implies maximality of all higher Rényi entropies and hence of the von Neumann entropy.

Since the density matrix $\rho_B$ is not necessarily normalized, its second Rényi entropy is defined by
\begin{equation}
    e^{-S_2(\rho_B)} = \frac{\text{Tr}(\rho_B^2)}{\text{Tr}(\rho_B)^2}
\end{equation}
where Tr is the trace over the boundary Hilbert space. The swap trick allows us to rewrite the traces and define $Z_1$ and $Z_0$ as:
\begin{align}
    Z_1 &\coloneqq \text{Tr}(\rho_B^2) = \text{Tr}(S_B\rho_B \otimes \rho_B)
    \label{Z_1}\\
    Z_0 &\coloneqq \text{Tr}(\rho_B)^2 = \text{Tr}(\rho_B \otimes \rho_B)
    \label{Z_0}
\end{align}
where $S_B$ is the swap operator acting on $ \rho_B \otimes \rho_B$. If we let $Z_0 = \avg{Z_0} + \delta Z_0$ and $Z_1 = \avg{Z_1} + \delta Z_1$, we can express $ \avg{S_2(\rho_B)}$ by 

\begin{align}
\begin{split}
    \avg{S_2(\rho_B)} &= - \avg{ \log \frac{\avg{Z_1} + \delta Z_1} { \avg{Z_0}+\delta Z_0}} \\
    &= -\log \frac{\avg{Z_1}}{\avg{Z_0}} + \sum_{k=1}^\infty \frac{(-1)^{k+1}}{k}\Big{(}\frac{\avg{\delta Z_0^k}}{\avg{Z_0^k}}-\frac{\avg{\delta Z_1^k}}{\avg{Z_1^k}}\Big{)}
\end{split}
\end{align}

within the series' radius of convergence. As shown in~\cite{HoloRTN}, when all the bond dimensions $E_l$ are large , the corrections are suppressed. Because of this result, we introduce a lower bound $J_{min}$ for nonzero spins appearing in Eq.~\eqref{linear_superposition}. The valid range of spins then becomes $\{0,J_{min},J_{min}+1/2,J_{min}+1,...\}$. For sufficiently large $J_{min}$, we can approximate the second Rényi entropy by
\begin{equation}
    \langle S_2(\rho_B) \rangle_\mu \approx \log \frac{\avg{Z_0}}{\avg{Z_1}}
    \label{S_2}
\end{equation}
To compute $\avg{Z_1}$, we need to substitute Eq.~\eqref{Psi(a_l)} and definition~\eqref{isometry} into definition~\eqref{Z_1}. Notice that the average $\avgn{-}$ only acts on vertex states, so we can define
\begin{equation}
    \rho \coloneqq \bigotimes_{l\in L}\big{(}\frac{1}{E_l}\sum_{a_l=0}^{E_l-1}\ket{a_l}\bra{a_l} \big{)}\bigotimes_{b\in B} \ket{l_b}\bra{l_b}
    \label{eqn:rho}
\end{equation}
and express $ \avg{Z_1}$ by a trace over internal links, node intertwiners, and boundary links
\begin{equation}
    \avg{Z_1} = \text{Tr}[(\rho \otimes \rho)S_B \bigotimes_{n\in N} \avgn{\ket{n}\bra{n} \otimes \ket{n}\bra{n}} ]
    \label{Z_1_rewrite}
\end{equation}
It was shown in~\cite{AvgResult} that 
\begin{equation}
    \avgn{\ket{n}\bra{n} \otimes \ket{n}\bra{n}} = \frac{\mathbbm{1}_n+ S_n}{D_n(D_n+1)}
    \label{vertex_avg}
\end{equation}
where $ \mathbbm{1}_n$ and $S_n$ are the identity operator and swap operator on $ \mathcal{H}_n \otimes \mathcal{H}_n$, respectively, and $D_n=dim(\mathcal{H}_n)$ is the dimension of the spin-network vertex space of node $n$. Substituting Eq.~\eqref{vertex_avg} into Eq.~\eqref{Z_1_rewrite} gives
\begin{equation}
    \avg{Z_1} = C^{-1} \text{Tr}[(\rho \otimes \rho)S_B \bigotimes_{n\in N}(\mathbbm{1}_n+ S_n) ]
    \label{1_n+S_n}
\end{equation}
where $C = \prod_n D_n(D_n+1)$. Since for any two nodes $n_1$ and  $n_2$, the following properties hold:
\begin{align}
    \begin{split}
        \mathbbm{1}_{n_1} \otimes \mathbbm{1}_{n_2} &= \mathbbm{1}_{\{n_1,n_2\}}\\
        S_{n_1} \otimes S_{n_2} &= S_{\{n_1,n_2\}}
    \end{split}
\end{align}
we can turn Eq.~\eqref{1_n+S_n} into a sum over the power set $2^N$ of $N$:
\begin{equation}
    \avg{Z_1} = C^{-1}\sum_{A \in 2^N}\text{Tr}[(\rho \otimes \rho)S_{B \cup A}]
\end{equation}
In random tensor network models, the calculation of this quantity is often performed~\cite{HoloRTN} via Ising-like variables by defining $s_n = +1$ if $n\not \in A$ and $s_n = -1$ if $n\in A$. Using this description, summing over the power set of $N$ is equivalent to summing over all possible spin configurations $\{s_n\}_{n\in N}$. We can then view $\avg{Z_1}$ as the partition function of the Ising spin system $\avg{Z_1} = \sum_{\{s_n\}}e^{-\mathcal{A}_1(\{s_n\})}$ and
\begin{equation}
    e^{-\mathcal{A}_1(\{s_n\})} = C^{-1}\text{Tr}[(\rho \otimes \rho)S_B \bigotimes_{s_n=-1}S_n]
    \label{action}
\end{equation} 
Recall that the trace in this expression is over internal links, node intertwiners, and boundary spins. If the link degrees of freedom were allowed to explore superpositions of spin representations dynamically, the swap operator $S_n$ acting on $\mathcal H_n\otimes\mathcal H_n$ would not factorize into a tensor product of swap operators on individual links and the intertwiner space. In the present setting, however, the Rényi entropy calculation is performed within the effective bulk subspace generated by the states $\ket{a_l}$, with the spin labels treated as fixed background data. As a result, the local Hilbert space at each node factorizes, and inside the trace in Eq.~\eqref{action} the swap operator can be written as a tensor product over links and the intertwiner space:
\begin{equation}
    S_n = \bigotimes_{i=0}^{|N|-1}S_n^i
\end{equation}
where $S_n^0$ acts on the two copies of the intertwiner space, and $S_n^i$ acts on two copies of link $i$ connecting to the node $n$, with a total of $|N|-1$ links. If $n$ is connected to a boundary node, $S_n$ contains an extra factor $S_n^\partial$ that acts on two copies of the boundary link space. Because the trace of tensor product is equal to product of trace, we could separate Eq.~\eqref{action} into three terms tracing over internal links, node intertwiners, and boundary spins respectively:
\begin{equation}
    e^{-\mathcal{A}_1(\{s_n\})} = C^{-1}\text{Tr}_L \big{[}\rho_L^{\otimes2} \bigotimes_{s_n=-1}\bigotimes_{i=1}^{V-1}S_n^i\big{]}\,\text{Tr}_{\mathscr{I}}\big{[}\bigotimes_{s_n=-1} S_n^0\big{]}\, \text{Tr}_B\big{[}\rho_\partial^{\otimes2}S_B \bigotimes_{s_n=-1}S_n^\partial\big{]}
    \label{factored_action}
\end{equation}
where $ \rho_L = \bigotimes_{l\in L}\big{(}\frac{1}{E_l}\sum_{a_l=0}^{E_l-1}\ket{a_l}\bra{a_l} \big{)} $ and $\rho_\partial = \bigotimes_{b\in B} \ket{l_b}\bra{l_b}$. Hence the total contribution, after adding a constant term, is
\begin{align}
\begin{split}
     \mathcal{A}_1(\{s_n\}) &= -\sum_{l\in L}\frac{s(l)}{2}(s_ms_n-1) + \sum_{l\in L}\frac{\log E_l}{4}(1-s_m)(1-s_n)\\&+\frac{1}{2}\sum_{b\in B} (s_{n_b}-1) \log{D} -\sum_{n\in N}\frac{1}{2}(s_n-1)\log{D_{\mathscr{I}_n}}+|B|\log{D}
\end{split}
\end{align}
where $e^{-s(l)} = \frac{1}{E_l^2} \sum_{a=0}^{E_l-1}(2a+1)d^{-a}$. A detailed derivation is given in appendix~\ref{app:ising}. $\avg{Z_0} = \text{Tr} (\rho_B \otimes \rho_B)$ can be computed in a similar way which results in an effective action $\mathcal{A}_0(\{s_n\})$. The only difference from $\avg{Z_1}$ is now we don't have $S_B$ in the trace, hence we get
\begin{equation}
    e^{-\mathcal{A}_0(\{s_n\})} = C^{-1}\text{Tr}_L \big{[}\rho_L^{\otimes2} \bigotimes_{s_n=-1}\bigotimes_{i=1}^{V-1}S_n^i\big{]}\,\text{Tr}_{\mathscr{I}}\big{[}\bigotimes_{s_n=-1} S_n^0\big{]}\, \text{Tr}_B\big{[}\rho_\partial^{\otimes2} \bigotimes_{s_n=-1}S_n^\partial\big{]}
\end{equation}
and adding the same constant term to $\mathcal{A}_0$ as in $\mathcal{A}_1$,
\begin{align}
\begin{split}
     \mathcal{A}_0(\{s_n\}) &= -\sum_{l\in L}\frac{s(l)}{2}(s_ms_n-1) + \sum_{l\in L}\frac{\log E_l}{4}(1-s_m)(1-s_n)\\&-\frac{1}{2}\sum_{b\in B} (s_{n_b}-1) \log{D} -\sum_{n\in N}\frac{1}{2}(s_n-1)\log{D_{\mathscr{I}_n}}
\end{split}
\end{align}
When the size of the graph $V$ gets sufficiently large, $\avg{Z_1} = \sum_{\{s_n\}}e^{-\mathcal{A}_1(\{s_n\})}$ and $\avg{Z_0} = \sum_{\{s_n\}}e^{-\mathcal{A}_0(\{s_n\})}$ can be approximated by the Ising spin configurations with the minimal actions, $\avg{Z_1} = e^{-\min_{\{s_n\}}\mathcal{A}_1(\{s_n\})}$ and $\avg{Z_0} = e^{-\min_{\{s_n\}}\mathcal{A}_0(\{s_n\})}$. In this case, using Eq.~\eqref{S_2}, we get

\begin{equation}
    \avg{S_2(\rho_B)} \approx \min_{\{s_n\}}\mathcal{A}_1 - \min_{\{s_n\}}\mathcal{A}_0
    \label{S_2_as_A} 
\end{equation}
Observe that $\mathcal{A}_0 = 0$ when $s_n=1$ for all $v \in N$. Furthermore, flipping any set of Ising spins to $-1$ will result in a non-negative change in each of the three terms in $\mathcal{A}_0$. Hence this is the configuration with minimal $\mathcal{A}_0$. Therefore, we can eliminate the second term in Eq.~\eqref{S_2_as_A}, and our task reduces to finding the minimal $\mathcal{A}_1$.
\\\\
When $s_n=1$ for all $n \in N$, $S_2 = \mathcal{A}_1 = |B| \log D$ which is the entropy of a maximally entangled state between bulk and boundary regions. This is because the bulk subspace always has higher dimension than the boundary subspace as the graph becomes larger, so a pure state in the entire Hilbert space could maximally have entanglement entropy equal to the log of the dimension of the smaller subspace, $|B| \log D$. Therefore, if this state results in $ \min_{\{s_n\}}\mathcal{A}_1$, $V^\dagger:\mathcal{H}_B\to\mathcal{H}_{\mathrm{bulk}}$ is an isometry.

\subsection{Isometry Condition with Quadrivalent Nodes}

Any spin configuration could be described by a subset $A \subseteq N$ such that $s_n=-1$ if and only if $n \in A$. The Ising action of this configuration is

\begin{equation}
     \mathcal{A}_1(A) = \sum_{l\in L \cap \partial A}s(l) + \sum_{l\in A} \log{E_l}- \sum_{b\in B\cap \partial A} \log{D} +\sum_{n\in A}\log{D_{\mathscr{I}_n}}+|B|\log{D}
\end{equation}

First, let $A = \{n\}$ where $n$ is connected to a boundary node. If the minimal allowed spin satisfies $J_{min} \gg 1$, we can approximate $s(l)$ by $2\log{E_l}$ where $d^{E_l-1}=2j(l)+1$. Let $j_1,j_2,j_3$ be the spins on the internal links of $n$, and $j_4=j$ be the spin on the boundary link of $n$. The Ising action then reads
\begin{equation}
     \mathcal{A}_1(\{n\}) = 2\log(E_1E_2E_3) - \log(2j_4+1) +\log{D_{\mathscr{I}_n}}+|B|\log{D}
\end{equation}
Therefore, the condition for $ \mathcal{A}_1(\{n\}) \ge |B|\log{D}$ becomes
\begin{equation}
    D_{\mathscr{I}_n} \ge \frac{2j_4+1}{(E_1E_2E_3)^2}
\end{equation}
This is a relatively weak constraint on the thickness of the intertwiner space. For a 4-valent node, the intertwiner dimension is 
\begin{equation}
    D_{\mathscr{I}_n} = \min\{j^1+j^2,j^3+j^4\} -\max\{|j^1-j^2|,|j^3-j^4|\}+1
\end{equation}

We may treat $j_4$ as the largest spin since it's the worst case scenario. The quantity $D_{\mathscr{I}_n}$ has a direct geometric interpretation. 
For a 4-valent node, choosing a recoupling channel amounts to choosing an intermediate spin. For any 
nondegenerate large-spin node, the overlap of the two Clebsch-Gordan intervals has macroscopic width, so $D_{\mathscr{I}_n}$ is of order $j_4$. The inequality therefore excludes only finely 
tuned boundary-layer configurations in which the local polyhedral shape space becomes anomalously thin. Geometrically, this occurs only when the spin data lie close to the boundary of the fusion polytope, for example when one face area is almost saturated by the sum of the others. Such configurations correspond to nearly degenerate polyhedra rather than to a generic bulk geometry.

Next, we consider arbitrary $A$ by adding nodes to $A$ one at a time. If we add a boundary node, what we computed tells us that this process increases the action. If we add a bulk node, all contributions are positive so we also increase the action. Therefore, the isometry condition is trivially satisfied for sufficiently large $J_{min}$. The above argument assumes that the links are spin eigenstates, but it is easy to see that introducing superposition of spins does not change the result. Since the boundary spin $j_4=j$ is fixed, when $J_{min}$ is large enough, even the smallest nonzero spin on each bulk link will still sum to be larger than $j_4$. We could thus obtain a lower bound of the change in action by treating the links as if they are spin eigenstates with the smallest spin in the linear combination, and this change in action will still be positive.

\section{Code Subspace}\label{sec:code_subspace}

\subsection{Setup and Definitions}

In the previous section we have shown that the bulk Hilbert space has larger dimension than the boundary Hilbert space. 
As a consequence, the bulk-to-boundary map $V:\{a_l\}\mapsto \ket{\Psi(\{a_l\})}$ cannot be injective when acting on the full space of bulk configurations. Nevertheless, in holographic constructions (particularly those inspired by tensor network models) one is typically not interested in the entire bulk Hilbert space, but rather in a restricted set of configurations describing perturbations around a fixed background structure. Here, we adopt this perspective in the context of spin networks and quantum geometries. Importantly, the notion of “classical geometry” used in this section should not be understood as a classical limit in the sense of quantum simplicial (lattice) geometry. Rather, it refers to a configuration with a \emph{well-defined combinatorial structure}, in which links are effectively present or absent and the underlying graph is sharply specified. This notion is directly analogous to the choice of a background tensor network on which small fluctuations are considered.

In this section, we make this notion precise by characterizing classical geometries in terms of the entanglement variables $\{a_l\}$ introduced earlier. Each such configuration determines a subspace $\mathcal{H}(\{a_l\})$ of the bulk Hilbert space. When the bulk-to-boundary map is restricted to this subspace, it is expected to inject into the boundary Hilbert space and to act as an isometry. In this case, $\mathcal{H}(\{a_l\})$ is isomorphically mapped to a subspace of the boundary Hilbert space, which we refer to as the \emph{code subspace}, due to the following reconstruction property.\\
For any region $A \subset B$ on the boundary, one can associate a corresponding bulk region following the standard holographic intuition developed in AdS/CFT and tensor-network models. In continuum holography, this association is formulated in terms of a minimal bulk surface $\gamma_A$ and the corresponding entanglement wedge. Here, we adopt the same conceptual framework, but implement it in a purely \emph{kinematical and discrete} setting adapted to spin networks. Concretely, for a fixed spin-network graph, we define $\gamma_A$ as a minimal cut in the bulk graph that separates the boundary region $A$ from its complement. This cut plays the role of a discrete minimal surface, in direct analogy with random tensor network constructions. The bulk region enclosed by $A$ and $\gamma_A$ will be denoted by $E_A$ and referred to as the \emph{entanglement wedge} associated with $A$.

We stress that, unlike in continuum AdS/CFT, spin networks do not define a spacetime geometry with causal structure or time evolution. They are instead interpreted as quantized spatial geometries, or equivalently as discretizations of a Cauchy surface. Accordingly, the entanglement wedge $E_A$ is defined here as a purely combinatorial and graph-theoretic notion, without reference to domains of dependence or Lorentzian dynamics. Nevertheless, this notion is sufficient to capture the structural features of holographic reconstruction familiar from AdS/CFT and tensor-network models.

Let $\hat{O}:\mathcal{H}_{\mathrm{bulk}}\to\mathcal{H}_{\mathrm{bulk}}$ be an operator satisfying the following two conditions:
\begin{itemize}
    \item[(1)] $\hat{O}$ acts nontrivially only within the code subspace
    $\mathcal{H}(\{a_l\})$, i.e.\ $\hat{O}\ket{\{a'_l\}}=\ket{\{a'_l\}}$
    for all $\ket{\{a'_l\}}\notin\mathcal{H}(\{a_l\})$;
    \item[(2)] $\hat{O}$ acts nontrivially only on bulk degrees of freedom associated with links
    entirely contained in $E_A$. That is, if one or both nodes attached to $l$ are not in $E_A$, then $ \hat{O}\ket{\{a'\}} = \hat{O}\ket{...a_{l'},a_l,...} = |...\hat{O}(a'_{l'}),a'_l,...\rangle$.
\end{itemize}

Under these assumptions, by analogy with random tensor networks~\cite{HoloRTN}, we expect $\hat{O}$ to be reconstructable by (isomorphically mapped to) an operator on $\mathcal{H}_A \subset \mathcal{H}_{B}$.
For each such operator $\hat{O}$, there might exist multiple boundary regions from which $\hat{O}$ can be reconstructed. In this sense, the reconstruction process acts like a quantum error correcting code and the corresponding code subspace is the image of $\mathcal{H}(\{a_l\})$ under $V$.

To construct the code subspace, we first need to define what a classical geometry is. Recall that the entanglement variable $a_l$ on each link $l$ takes values between $0$ and $E_l-1$, with $E_l$ very large. A geometry $\{a_l\}$ is defined to be classical if for all $l \in L$, $\frac{a_l}{E_l-1} \in [0,1]$ is either $0$ or $O(1)$. Fix $\Lambda \ll E_l$, and the code subspace $\mathcal{H}(\{a_l\})$ is defined to be the subspace generated by all the states $\{a'_l\}$ such that for all $l \in L$,

\begin{equation}\label{eq:axy‐range}
a'_{l}\in
\begin{cases}
[a_l - \Lambda,\;a_l + \Lambda], & \text{if } a_l \not=0,\\
[0,\;2\Lambda],                   & \text{if } a_l=0~.
\end{cases}
\end{equation}

Let $W$ be the restriction of the bulk-to-boundary map $V$ to $\mathcal{H}(\{a_l\})$, and define the boundary code subspace as
\begin{equation}
    \mathcal{H}_{\mathrm{code}} \coloneqq \mathrm{im}\, W \subseteq \mathcal{H}_B .
\end{equation}
Note that, throughout this section, the intertwiners are not treated as arbitrary independent local degrees of freedom. Rather, their allowed fluctuations are constrained by compatibility with the link data, and after averaging they contribute through the
corresponding effective entropic factors. 
In particular, the quantities $D_{\mathscr I_n}$ that will appear in the isometry condition measure the dimension of the space of intertwiners compatible with a given configuration of link variables within the code subspace, but do not correspond to additional local Hilbert-space factors.

A concrete realization of intertwiner fluctuations compatible with the above definition of the code subspace is presented in Appendix~\ref{app:intertwiners}. 
The details of this construction will not be needed in the following.

\subsection{Subspace Isometry Condition}

We now examine the conditions under which the bulk-to-boundary map $W:\mathcal{H}(\{a_l\})\to\mathcal{H}_{\mathrm{code}}$ is an isometry. Recall that $W$ is defined as the restriction of the map $V:\mathcal{H}_{\mathrm{bulk}}\to\mathcal{H}_B$ to the subspace
$\mathcal{H}(\{a_l\})$.

Its adjoint is therefore not simply the restriction of $V^\dagger$ to $\mathcal{H}_{\mathrm{code}}$, but rather
\begin{equation}
W^\dagger = P_{\mathcal{H}(\{a_l\})} \circ V^\dagger\big|_{\mathcal{H}_{\mathrm{code}}},
\end{equation}
where $P_{\mathcal{H}(\{a_l\})}$ denotes the projector onto the bulk subspace $\mathcal{H}(\{a_l\})$. As a consequence, $W^\dagger$ is not automatically an isometry.

\begin{figure}
    \centering
    \[\begin{tikzcd}[sep=large]
	{\mathcal{H}_{bulk}} && {\mathcal{H}_B} \\
	\\
	{\mathcal{H}(\{a_l\})} && {\mathcal{H}_{code}}
	\arrow["V", two heads, from=1-1, to=1-3]
	\arrow["{V^\dagger}", shift left=3, hook', from=1-3, to=1-1]
	\arrow[hook, from=3-1, to=1-1]
	\arrow["W", from=3-1, to=3-3]
	\arrow[hook', from=3-3, to=1-3]
	\arrow["{W^\dagger}", shift left=3, from=3-3, to=3-1]
\end{tikzcd}\]
    \caption{Maps between various Hilbert spaces we are considering. The bulk-to-boundary map $V$, defined in section~\ref{sec:framework}, is not an isometry since it is not injective. In section~\ref{sec:Btb_iso}, we showed that when $J_{min}$ is sufficiently large, the boundary-to-bulk map $V^\dagger$ is an isometry. $W$ is the restriction of $V$ on $\mathcal{H}(\{a_l\})$. In section~\ref{sec:code_subspace}, we showed that when $J_{min}$ is sufficiently large, both $W$ and $W^\dagger$ are isometries and they are inverses of each other. }
    \label{fig:hilbert_maps}
\end{figure}
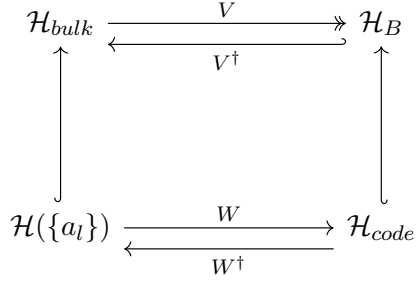

A convenient sufficient criterion for $W$ to be an isometry consists of the following two conditions:
\begin{enumerate}
    \item $\dim \mathcal{H}(\{a_l\}) = \dim \mathcal{H}_{\mathrm{code}}$;
    \item $WW^\dagger = \mathbbm{1}_{\mathcal{H}_{\mathrm{code}}}$, i.e. $W^\dagger$ is an isometry.
\end{enumerate}
Indeed, since $W$ is surjective by definition, condition (1) implies that $W$ is injective and hence a linear isomorphism. Condition (2) then implies that $W^\dagger = W^{-1}$, from which the isometry condition $W^\dagger W = \mathbbm{1}_{\mathcal{H}(\{a_l\})}$
follows immediately. The relations between the various Hilbert spaces and maps introduced above are summarized in Fig.~\ref{fig:hilbert_maps}.

To show (2), we employ the same technique we used in section~\ref{sec:Btb_iso}, which is to show that
\begin{equation}
    \rho_B=\sum_{\{a'_l\}\in \mathcal{H}(\{a_l\})} \ket{\Psi(\{a'_l\})}\bra{\Psi(\{a'_l\})} = \mathbbm{1}_{\mathcal H_{\mathrm{code}}}
    \label{eqn:code_subspace_rho}
\end{equation}
The calculation is very similar to that in the previous section, except we need to replace the link state $\rho_L$ appearing in Eq.~\eqref{factored_action} with
\begin{equation}
    \rho_L = \bigotimes_{l\in L}\rho_l
\end{equation}
where
\begin{equation}\label{rho_l}
\rho_{l}=
\begin{cases}
\frac{1}{2\Lambda+1}\sum_{i=-\Lambda}^\Lambda \ket{a_l+i}\bra{a_l+i}, & \text{if } a_l \not=0,\\
\frac{1}{2\Lambda+1}\sum_{i=0}^{2\Lambda} \ket{i}\bra{i},                   & \text{if } a_l=0~.
\end{cases}
\end{equation}
Let $L_0$ be the set of all links with $a_l=0$ (classically disconnected), and $L_1 = L - L_0$. The resulting actions are
\begin{align}
    \begin{split}
        \mathcal{A}_1(\{s_n\}) &= -\sum_{l\in L_0}\frac{s_0(l)}{2}(s_ms_n-1)-\sum_{l\in L_1}\frac{s_1(l)}{2}(s_ms_n-1)+\sum_{l\in L}\frac{\log(2\Lambda+1)}{4}(1-s_m)(1-s_n)\\ &+ \frac{1}{2}\sum_{b\in B} (s_{n_b}-1) \log{D} -\sum_{n\in N}\frac{1}{2}(s_n-1)\log{D_{\mathscr{I}_n}}+|B|\log{D}
    \end{split}
\end{align}
and
\begin{align}
    \begin{split}
        \mathcal{A}_0(\{s_n\}) &= -\sum_{l\in L_0}\frac{s_0(l)}{2}(s_ms_n-1)-\sum_{l\in L_1}\frac{s_1(l)}{2}(s_ms_n-1)+\sum_{l\in L}\frac{\log(2\Lambda+1)}{4}(1-s_m)(1-s_n)\\ &- \frac{1}{2}\sum_{b\in B} (s_{n_b}-1) \log{D} -\sum_{n\in N}\frac{1}{2}(s_n-1)\log{D_{\mathscr{I}_n}}
    \end{split}
\end{align}
The minimum of $\mathcal{A}_0 $ occurs when $s_n=1$ for all $n\in N$, with $ \mathcal{A}_0(\{s_n=1\})$ = 0. Therefore, the averaged second Rényi entropy is given by
\begin{equation}
    \avg{S_2(\rho_B)} \approx \min_{\{s_n\}}\mathcal{A}_1
    \label{S_2_as_A_1} 
\end{equation}
When $s_n=1$ for all $n\in N$, 
\begin{equation}
  \mathcal{A}_1(\{s_n=1\}) = |B| \log{D}  
\end{equation}
When $s_n=-1$ for all $n\in N$,
\begin{align}
    \begin{split}
        \mathcal{A}_1(\{s_n=-1\}) &= |L|\log{(2\Lambda+1)}-|B| \log{D}+\sum_{n\in N}\log{D_{\mathscr{I}_n}} +|B| \log{D}\\
        &=\frac{1}{2}|N|(|N|-1)\log{(2\Lambda+1)}+\log{D_{\mathscr{I}}}\\
        &=|N|(|N|-1)\log{(2\Lambda+1)}
        \label{eq:A_1min}
    \end{split}
\end{align}
where the computation of $D_{\mathscr{I}}$ is performed in Appendix~\ref{app:intertwiners}, Eq.~\eqref{eq:intertwiner-subspace-dimension}. For any fixed graph, we choose $D$ large enough such that $\mathcal{A}_1(\{s_n=-1\}) < \mathcal{A}_1(\{s_n=1\})$. 

If $\min_{\{s_n\}}\mathcal{A}_1=\mathcal{A}_1(\{s_n=-1\})$, the entanglement entropy of $\rho_B$ is the log of the dimension of $\mathcal{H}(\{a_l\})$. The maximal possible entanglement entropy of $\rho_B$ is $\log \dim{\mathcal{H}}_{code}$, as $\dim{\mathcal{H}}_{code} \le \dim \mathcal{H}(\{a_l\})$ since $\mathcal{H}_{code}$ was defined to be the image of $\mathcal{H}(\{a_l\})$ under $W$. Therefore, we are forced to have $\dim{\mathcal{H}}_{code} = \dim \mathcal{H}(\{a_l\})$. Thus, conditions (1) and (2) stated at the beginning of this section are both satisfied, and hence $W$ is an isometry.

It remains to check if $\mathcal{A}_1$ takes minimal value for all spin down configuration. The action could be described by a subset $A \subseteq N$ such that $s_n=-1$ if and only if $n \in A$. The Ising action of this configuration is

\begin{align}
\begin{split}
     \mathcal{A}_1(A) &= \sum_{l\in L_0 \cap \partial A}s_0(l)+ \sum_{l\in L_1 \cap \partial A}s_1(l)+\sum_{l\in A}\log(2\Lambda+1)\\&- \sum_{b\in B\cap \partial A} \log{D} +\sum_{n\in A}\log{D_{\mathscr{I}_n}}+|B|\log{D}
     \end{split}
\end{align}
where under the conditions $1 \ll\Lambda\ll a_l$, $s_0(l) \approx 2\log(2\Lambda+1)$ and $s_1(l) \approx (a_{nm}-\Lambda)\log d+s_0$.
We already saw that if $A=N$, then $ \mathcal{A}_1(N) = N(N-1)\log{(2\Lambda+1)}$. Now we flip the Ising spin $s_n$ at node $n$. First suppose that $n$ is a bulk node. Let the adjacent nonzero classical entanglement entropy variables be $a_1,a_2,a_3,a_4$.
\begin{align}
    \begin{split}
      \mathcal{A}_1(N-\{n\})-\mathcal{A}_1(N) &= (|N|-5)s_0+(a_1+a_2+a_3+a_4-4\Lambda)\log d + 4s_0 - \frac{1}{2}(|N|-1)s_0- \log D_{\mathscr{I}_n}\\
      &=\frac{1}{2}(|N|-1)s_0+(a_1+a_2+a_3+a_4-4\Lambda)\log d - \log D_{\mathscr{I}_n}
    \end{split}
\end{align}
Note that by assumption, $a_i$ is of the same order as $E_i$, and $ \log j_i = (E_i-1)\log d$. So in the large $J_{min}$ limit the sum of the latter two terms is always positive. Next, suppose $n$ is a node connected to a boundary node. Since the intertwiner space is the invariant subspace $\mathrm{Inv}_{SU(2)}\Bigl(V_{j_b}\otimes\,\bigotimes_{m\neq n}V_{\,a_{nm}}\Bigr)$, where $j_b$ is the spin on the boundary link, its dimension cannot exceed the dimension of $V_{j_b} $, which is $D = 2j_b+1$. We also only have three nonzero classical entanglement entropy variables $a_1,a_2,a_3$. So we get
\begin{align}
    \begin{split}
      \mathcal{A}_1(N-\{n\})-\mathcal{A}_1(N) &= \frac{1}{2}(|N|-1)s_0+(a_1+a_2+a_3-3\Lambda)\log d +\log D-\log D_{\mathscr{I}_n}\\
      &\ge\frac{1}{2}(|N|-1)s_0+(a_1+a_2+a_3-3\Lambda)\log d\\
      &>0
    \end{split}
\end{align}
and the change to the action is positive. By similar analysis, further removing points from $A$ always results in an increase in the action.
Thus, for any graph $(N,L,B)$, for sufficiently large $J_{min}$ relative to $\Lambda$, the map $W:\mathcal{H}(\{a_l\}) \to \mathcal{H}_{code}$ obtained by restricting $V: \{a_l\} \mapsto \ket{\Psi(\{a_l\})}$ is both an isomorphism and an isometry.

\subsection{Overlap Between States}
\label{sec:overlap}

Since the bulk-to-boundary map $V: \mathcal{H}_{bulk} \to \mathcal{H}_B$ is not injective, we expect the code subspaces $\mathcal{H}(\{a_l\})$ and $\mathcal{H}(\{b_l\})$ to have nontrivial overlap even for completely distinct (i.e. different at every link) classical geometries $\{a_l\}$ and $\{b_l\}$. In this section, we compute the first and second moments of this overlap $C_{ab}$ which is defined as
\begin{equation}
    C_{ab}\coloneqq \ip{\Psi(\{a_l\})}{\Psi(\{b_l\})}
\end{equation}
Using definition~\ref{bdy_state} and factorization property of trace, we get
\begin{align}
    \begin{split}
        C_{ab} &=\text{Tr}_{L,\mathscr{I},B}[\bigotimes_{l\in L}\ket{a_l}\bra{b_l}\bigotimes_{n\in N}\ket{n}\bra{n}\bigotimes_{b\in B}\ket{l_b}\bra{l_b}]\,\\
        &= \text{Tr}_{L,\mathscr{I}}[\bigotimes_{l\in L}\ket{a_l}\bra{b_l}\bigotimes_{n\in N}\ket{n}\bra{n}]\,\prod_{b\in B}\ip{l_b}{l_b}^2\\
    &= \text{Tr}_{L,\mathscr{I}}[\bigotimes_{l\in L}\ket{a_l}\bra{b_l}\bigotimes_{n\in N}\ket{n}\bra{n}]
    \end{split}
\end{align}
since $\ket{l_b}$ is normalized. The expected value of the overlap is thus
\begin{equation}
    \avg{C_{ab}} = \text{Tr}_{L,\mathscr{I}}[\bigotimes_{l\in L}\ket{a_l}\bra{b_l}\bigotimes_{n\in N}\avgn{\ket{n}\bra{n}}]
\end{equation}
The second average can be evaluated by using the identity
\begin{equation}
    \avgn{\ket{n}\bra{n}} = \frac{\mathbbm{1}_n}{D_n}
\end{equation}
The identity operator acts trivially on the first part, so the result becomes
\begin{equation}
    \avg{C_{ab}} = \frac{1}{D_\mathscr{I}}\text{Tr}_{L,\mathscr{I}}[\bigotimes_{l\in L}\ket{a_l}\bra{b_l}]
\end{equation}
where $ D_\mathscr{I} = \prod_{n\in N}D_n$. Since the expression in the trace does not depend on the intertwiner anymore, the trace over $\mathscr{I}$ just produces a multiplicative factor of $D_{\mathscr{I}}$ which exactly cancels the prefactor. The remaining trace over link state gives
\begin{align}
    \begin{split}
        \text{Tr}_L[\bigotimes_{l\in L}\ket{a_l}\bra{b_l}] &= \prod_{l\in L}\sum_{i=0}^{E_l-1}\ip{i}{a_l}\ip{b_l}{i}\\
        &=\prod_{l\in L}\sum_{i=0}^{E_l-1}\delta_{i,a_l}\delta_{b_l,i}\\
        &=\prod_{l\in L}\delta_{a_l,b_l}\\
        &\eqqcolon\delta_{ab}
        \end{split}
\end{align}
Thus we arrive at the simple result
\begin{equation}
    \avg{C_{ab}} = \delta_{ab}
\end{equation}
On average, there is no overlap between the images of two distinct classical states. This alone, however, tells us nothing about the overlap given any particular choice of intertwiners. To obtain a typical overlap, we need to compute the amplitude of the second moment
\begin{equation}
    \avg{|C_{ab}|^2}=\text{Tr}_{L,\mathscr{I},B}[\bigotimes_{l\in L}\ket{a_l}\bra{b_l}\otimes \ket{b_l}\bra{a_l}\bigotimes_{n\in N}\avgn{\ket{n}\bra{n}\otimes \ket{n}\bra{n}}\bigotimes_{b\in B}\ket{l_b}\bra{l_b}\otimes\ket{l_b}\bra{l_b}]
\end{equation}
After using a series of tricks shown in Appendix~\ref{app:overlap}, we arrive at
\begin{equation}
    \avg{|C_{ab}|^2}=\tilde{C}\sum_{A\in2^{N}}\text{Tr}_{L_A}[\rho_A^a\rho_A^b]\,D^{ |A\cap B|-|B|}
\end{equation}
This is again a sum over the power set of $N$ and $\rho^a = \bigotimes_{l\in L}\ket{a_l}\bra{a_l}$ and $\rho^b=\bigotimes_{l\in L}\ket{b_l}\bra{b_l}$ are the density matrices corresponding to $\{a_l\}$ and $\{b_l\}$ states, respectively. The reduced density matrix $\rho^a_{A} \coloneqq \text{Tr}_{L-L_A}[\rho^a]$ is defined by taking the partial trace over links not completely in $A$. The Cauchy-Schwarz inequality (with Hilbert-Schmidt inner product) gives an upper bound of the trace in terms of the second Rényi entropies $S_a^{(2)}(A)$ and $S_b^{(2)}(A)$ of the states $\rho_A^a$ and $\rho_A^b$,
\begin{equation}
    \text{Tr}_{L_A}[\rho_A^a\rho_A^b] \le \sqrt{\text{Tr}_{L_A}[(\rho_A^a)^2]\text{Tr}_{L_A}[(\rho_A^b)^2]} = e^{-\frac{1}{2}\big{(}S_a^{(2)}(A)+S_b^{(2)}(A)\big{)}}
\end{equation}
Furthermore, the trace over region $A$ is $0$ unless $\{a_l\}$ and $\{b_l\}$ agree in $A$ since we can factor the trace as the product on individual links. Therefore, instead of summing over all sets $A$ in the power set $2^N$, we only need to sum over its subset $\mathcal{C} \subset 2^N$ consisting of regions in which $\{a_l\}$ and $\{b_l\}$ completely agree. We thus obtain the bound
\begin{equation}
    \avg{|C_{ab}|^2} \le \tilde{C}\sum_{A\in \mathcal{C}}e^{-\frac{1}{2}\big{(}S_a^{(2)}(A)+S_b^{(2)}(A)\big{)}-\log D (|B| - |A\cap B|)}
    \label{upper_bound}
\end{equation}
This inequality allows us to examine the overlap between classically distinct states. First, consider the case where $a_l$ and $b_l$ are different on every link. This gives $\mathcal{C} = \{\emptyset\}$ so the sum has only one term
\begin{equation}
    \avg{|C_{ab}|^2} \le \tilde{C}D^{-|B|}
\end{equation}
Up to normalization, the fluctuation is bounded by the dimension of the boundary Hilbert space, which is also the expected overlap between two uniformly randomly chosen states.

Next, consider two spaces that differ only in a compact region $K$. For instance, one space has a Schwarzschild black hole while the other has a spherically symmetric regular piece of  matter with equal mass. Outside a finite ball $K$, the metrics are identical. Using our link variable description, $\{a_l\}$ and $\{b_l\}$ are equal in $K^c$ and distinct in $K$. From Eq.~\eqref{upper_bound}, we see that the most dominant summand $A \subset K^c$ is the one that minimizes $S_a^{(2)}(A),S_b^{(2)}(A)$, and maximizes $|A\cap B|$. Since our setup is spherically symmetric, each link contributes the same factor to the second Rényi entropy and hence it follows an area law
\begin{equation}
    S_{a,b}^{(2)}(A) = s_0|\partial A|
\end{equation}
where $ \partial A$ is the inner boundary of $A$. In particular, choosing $A$ to be larger reduces $|\partial A|$. Therefore, the dominant contribution is $A=K^c$ with $|A\cap B| = |B|$ and $ S_{a,b}^{(2)}(A) = s_0|\partial K|$. In a more general, non spherically symmetric setting, $A$ is still chosen to minimize the area $|\partial A|$. When the graph is large, we can ignore subleading terms in the sum and the bound becomes
\begin{equation}
    \avg{|C_{ab}|^2} \le e^{-s_0|\partial K|}
\end{equation}
As in the previous case, this is, up to the prefactor, the typical overlap of two uniformly random states, now in a Hilbert space of dimension $e^{s_0|\partial K|}$, set by the area of the boundary of the region where the two configurations differ. If holography holds, this is precisely the dimension of the Hilbert space associated with $K$.
In our black hole example, the regular matter is less dense than the black hole so it occupies a larger region, which means $|\partial K|$ is larger than the black hole horizon area $A_{BH}$. Hence, the overlap between these two geometries is bounded by the black hole entropy
\begin{equation}
     \avg{|C_{ab}|^2} \le e^{-S_{BH}}
\end{equation}
This means that any black hole microstate and any state describing the regular matter with the same global charges overlap as two random states in the holographic Hilbert space of $K$. Although the two configurations are asymptotically identical, their overlap is exponentially suppressed and controlled by the area of the region in which they differ rather than by its volume.

\section{Discussion}\label{sec:discussion}

In this work we have introduced a new way of endowing spin-network states with a controlled and flexible entanglement structure. The key ingredient is a generalization of the standard gluing prescription for spin networks, in which internal links are no longer restricted to the unique maximally entangled, gauge-invariant singlet state. Instead, each link is allowed to carry a discrete and tunable amount of entanglement between its endpoints, parametrized by an effective link variable $a_l$. This variable interpolates between product states and the fully entangled singlet, with only the latter preserving the diagonal $SU(2)$ gauge invariance at the link. In this way, internal links can carry nontrivial total spin, and the entanglement structure of the spin network becomes an explicit and adjustable part of the description, closely paralleling the role of bond variables in tensor-network constructions.

Within this enriched framework, the contraction of spin-network data naturally defines a bulk-to-boundary linear map whose input is the collection of link entanglement variables $\{a_l\}$ and whose output is a quantum state in the boundary Hilbert space. The vertex intertwiners do not appear as degrees of freedom encoded in the boundary state; rather, they specify the map itself, determining how bulk information is transferred to the boundary. This viewpoint motivates averaging over intertwiners with the natural Haar measure, which probes the typical properties of the resulting family of bulk-to-boundary maps. By analyzing the averaged Rényi entropy of the induced boundary state, we showed that for sufficiently large bulk spins and suitable graph structure the map preserves inner products in expectation value, providing a precise realization of holographic behavior directly within the spin-network Hilbert space.

Focusing on configurations in which the underlying graph has a fixed combinatorial structure—namely, configurations where links are distinguished by reference values of entanglement that are either close to maximal or vanishing—naturally selects a finite-dimensional code subspace of the bulk Hilbert space. Small fluctuations are described by allowing the entanglement on all links to vary within controlled windows around these reference values, while keeping the combinatorial structure of the graph fixed. In this regime, the restricted bulk-to-boundary map is both injective and an isometry up to corrections that vanish in the large-spin limit, with its adjoint providing the inverse map on the code subspace. This yields a fully discrete realization of a quantum error–correcting structure formulated entirely in terms of spin-network degrees of freedom.

Finally, we studied the overlap between boundary states associated with different bulk configurations. Averaging again over the vertex intertwiners, we found that distinct configurations are orthogonal on average, while the typical magnitude of their overlap is controlled by entropic properties of the region where the two configurations differ. When two configurations coincide outside a finite subset of links, the overlap is exponentially suppressed in terms of the size of the corresponding interface, providing a quantitative measure of the distinguishability of spin-network configurations encoded in boundary states. 
\smallskip

Although the overall logic of this construction resembles that of random tensor network models~\cite{HoloRTN,Qi:2017ohu} and, more generally, the quantum error-correcting description of holography~\cite{Almheiri:2014lwa,Harlow:2016vwg}, there are important new features. In the RTN setup of~\cite{Qi:2017ohu} the link variables fluctuate around classical values while the node tensors are independent of them; here the nodes are $SU(2)$ intertwiners, and the closure constraint~\cite{Livine:2007vk,Freidel:2007py} ties them to the incident link data. The code subspace therefore also includes compatible shape fluctuations around a Livine--Speziale coherent intertwiner (Appendix~\ref{app:intertwiners}). Moreover, the link spins carry a direct geometrical interpretation as areas of faces in a Regge-like discretization.

 Among existing holographic quantum error-correcting code models, the hyperbolic pentagon (HaPPY) code of Pastawski \emph{et al.}~\cite{Pastawski:2015qua} offers the most natural comparison: like our construction, it realizes an explicit bulk--to--boundary isometry on a fixed graph; in addition, it reproduces an RT-type minimal-surface formula for boundary entropies and implements entanglement wedge reconstruction, features that in the present work are only discussed qualitatively. A complementary construction, also motivated by holographic codes such as HaPPY, is provided by the hyperinvariant spin-network states of Otto \emph{et al.}~\cite{Otto:2026jhep}, which incorporate local $SU(2)$ invariance into hyperinvariant tensor networks. Their construction uses explicit tensors satisfying hyperinvariant isometry conditions; here we vary the entanglement on links and study the resulting maps through intertwiner averaging. They also prove that local $SU(2)$ invariance excludes holographic codes whose logical index transforms covariantly under the gauge group; our construction avoids this no-go, since the bulk data $\{a_l\}$ do not carry a representation of $SU(2)$ and the isometry holds only on average, in a large-spin regime. The closest tensor-network model with gauge symmetry is the gauged random tensor network of Dong, McBride and Weng~\cite{Dong:2024jhep}, where a gauge-invariant bulk state is fed into a standard random tensor network, yielding a non-trivial area operator. Here gauge invariance is instead imposed directly on the node tensors and the spins are fixed, so that the state dependence of the entanglement across a cut enters only through the link variables $a_l$.

Another potentially interesting application of our result is to pass from coarse spin networks to finer ones, i.e.\ embed a given state into a larger graph while preserving semiclassical continuity. In RTN holography, one often merges or splits tensors in a MERA network~\cite{Vidal:2007hda}. Analogously, one could imagine refining our complete‐graph code by splitting each tetrahedral node into four smaller tetrahedra (or merging adjacent tetrahedra into one higher‐valent node). The key question is how the code subspace changes: does its dimension remain large enough to maintain isometry? How do the Ising‐action weights flow under such moves? Answering these questions would amount to a discrete renormalization group on spin networks~\cite{Dittrich:2014wda}, and one that automatically respects holographic error correction. For instance, it may happen that merging two adjacent nodes into one higher‐valent node shifts the closure condition and can push certain spin assignments outside the small‐fluctuation window. Thus any refinement scheme must carefully adjust background spins or windows to have control over the evolution of the code subspace.\smallskip

We conclude by outlining a few open directions suggested by our construction.
Our analysis has been entirely kinematical and formulated on a fixed complete graph, where the code subspace and the associated bulk-to-boundary isometry can be sharply characterized. A natural next step is therefore to ask whether this structure is stable once dynamical amplitudes are taken into account. In particular, it would be interesting to investigate how spin-foam dynamics act on the restricted sector of states defining the code subspace, and whether the latter is preserved - at least approximately - under evolution. As a concrete test case, one may consider the EPRL/FK amplitude on a single 4-simplex dual to five spin-network nodes~\cite{Engle:2007wy,Freidel:2007py}, with boundary data chosen to match the background configuration and restricted to small-window fluctuations. One may then ask whether the resulting amplitude induces a map that acts within the code subspace and preserves the bulk-to-boundary isometry in an appropriate sense. Alternatively to spin-foam amplitudes, the stability of the code subspace could be tested under canonical dynamics, e.g.\ a Hamiltonian constraint. It would also be interesting to go beyond the large-spin regime and Livine--Speziale intertwiners, for instance using coherent states peaked also on the extrinsic geometry, which involve superpositions of spins. We leave a systematic investigation of these questions for future work.

The interpretation of link-enrichment parameters as edge-mode–like degrees of freedom suggests several natural extensions. In particular, it would be interesting to investigate whether these additional variables can be endowed with an effective dynamics, analogously to boundary degrees of freedom in gauge theories, and to clarify their role in gluing procedures and coarse-graining schemes in LQG and GFT. Moreover, the present analysis indicates that code-subspace–like structures and bulk-to-boundary isometries naturally emerge only in restricted sectors of the spin-network Hilbert space, typically associated with fixed graphs or specific link configurations. An important open direction is therefore to understand whether similar structures can arise for more general superpositions of spin networks, possibly after suitable coarse-graining or in appropriate semiclassical limits. Finally, the presence of additional physical degrees of freedom on links may offer new tools to address the reconstruction of graph structure in the GFT Fock-space, as by endowing links with relational data one may gain further control over combinatorial and permutation symmetries.

\acknowledgments

MQ would like to express his gratitude to Don Marolf for agreeing to supervise this project and for his ongoing support and guidance. MQ also thanks the UCSB College of Creative Studies and Tengiz Bibilashvili for creating a helpful environment. The authors thank Daniele Oriti for many helpful and inspiring discussions, and Simon Langenscheidt for comments on an earlier version of the manuscript. EC's participation in this project was made possible by a DeBenedictis Postdoctoral Fellowship and through the support of the ID\# 62312 grant from the John Templeton Foundation, as part of the \href{https://www.templeton.org/grant/the-quantum-information-structure-of-spacetime-qiss-second-phase}{``Quantum Information Structure of Spacetime'' Project (QISS)}. The opinions expressed in this project/publication are those of the authors and do not necessarily reflect the views of the John Templeton Foundation. EC was also supported by the ATRAE project PR28/23 ATR2023-145735 of the Agencia Estatal de Investigación (Spain), and is currently a Banting Postdoctoral Researcher at the University of Waterloo, supported by the Banting Postdoctoral Fellowships program in partnership with the Natural Sciences and Engineering Research Council of Canada (NSERC).

\appendix

\section{Gluing spin network vertices}
\label{app:gluing}

In this appendix we give the derivation of the gluing operation used in the main text. The goal is to show explicitly how two initially disjoint spin network vertices can be connected along a pair of open legs. At the level of group variables, the operation is implemented by averaging over a common left action of $SU(2)$ on the two legs to be glued, and we show that the resulting wavefunction depends only on the composed holonomy. In the
representation basis, this gluing corresponds to contracting the appropriate magnetic indices with the unique spin-$0$ (bivalent) intertwiner.\medskip

Let $n_1$ and $n_2$ be two spin network vertices with valences $k_1$ and $k_2$, respectively, and suppose we want to glue the $p$-th open leg of $n_1$ to the $q$-th open leg of $n_2$, thereby creating a new internal link of the graph. Let $\Psi_1(g^1,\ldots,g^{k_1})$ and $\Psi_2(h^1,\ldots,h^{k_2})$ be the corresponding $L^2$ wavefunctions. We define the glued wavefunction $\Psi$ by averaging over the common left multiplication:
\begin{equation}
    \Psi(g^1,...g^{k_1},h^1,...,h^{k_2}) \coloneq \int dx\; \Psi_1(g^1,...g^p x,...,g^{k_1}) \Psi_2(h^1,...,h^q x,...,h^{k_2})
    \label{glue}
\end{equation}
We can explicitly check that $\Psi$ only depends on $g^p$ and $h^q$ through the product $ g^p (h^q)^{-1}$. Consider the transformation
\[
g^p \to g^p y,
\qquad
h^q \to h^q y,
\]
which leaves the product $ g^p (h^q)^{-1}$ invariant. Then
\begin{align*}
    \begin{split}
     \Psi(g^1,...g^p y,...,g^{k_1},h^1,...,h^q y,...,h^{k_2}) &=    \int dx \;\Psi_1(g^1,...g^p yx,...,g^{k_1}) \Psi_2(h^1,...,h^q yx,...,h^{k_2})\\
     &= \int d(y^{-1}u) \;\Psi_1(g^1,...g^p u,...,g^{k_1}) \Psi_2(h^1,...,h^q u,...,h^{k_2})\\
     &= \int du \;\Psi_1(g^1,...g^p u,...,g^{k_1}) \Psi_2(h^1,...,h^q u,...,h^{k_2})\\
     &= \Psi(g^1,...g^p ,...,g^{k_1},h^1,...,h^q ,...,h^{k_2})
    \end{split}
\end{align*}
where the second equality uses the substitution $u = yx$, and the third equality uses left invariance of the Haar measure. Thus, the new function $\Psi$ depends on $k_1 + k_2 - 1$ independent variables. In the end, when all open legs in the graph are glued, the resulting spin network wavefunction depends on $L$ variables, one per link. \medskip

Returning for simplicity to the example of two four-valent vertices, we now show how the gluing works in the representation basis:
\begin{equation}
\begin{split}
\int dh \Psi_1(g^1,\dots,g^4h) &\Psi_2(q^1,\dots,q^4h)\\&= {\Psi_1}^{\vec{j}}_{\vec{m}\iota}{\Psi_2}^{\vec{j}'}_{\vec{m}'\iota'}\int dh \chi^{\vec{j}}_{\vec{m},\iota}(g^1,\dots,g^4h) \chi^{\vec{j}'}_{\vec{m}',\iota'}(q^1,\dots,q^4h)\\&= {\Psi_1}^{\vec{j}}_{\vec{m}\iota}{\Psi_2}^{\vec{j}'}_{\vec{m}'\iota'}\chi^{\vec{j}}_{m_1m_2m_3k,\iota}(\vec{g}) \chi^{\vec{j}'}_{m_1'm_2'm_3'k',\iota'}(\vec{q})\int dh \wig{j}{km_4}{h}\wig{j'}{k'm_4'}{h}\\&= \left({\Psi_1}^{\vec{j}}_{\vec{m}\iota}{\Psi_2}^{\vec{j}'}_{\vec{m}'\iota'}T_{m_4m_4'}\right)\left(\chi^{\vec{j}}_{m_1m_2m_3k,\iota}(\vec{g}) \chi^{\vec{j}}_{m_1'm_2'm_3'k',\iota'}(\vec{q})T_{kk'}\right)
\label{gluing_app}
\end{split}
\end{equation}
where we used
\begin{equation}\label{wig}
 \int dh \wig{j}{kn}{h}\wig{j'}{k'n'}{h}=\delta_{jj'}T_{kk'}T_{nn'}
\end{equation}
with $T_{kk'}\in V^{j}\otimes V^{j}$ the bivalent intertwiner 
\begin{equation}\label{bivalent_app}
T_{kk'}\coloneqq \frac{(-1)^{j+k}}{\sqrt{2j+1}}\delta_{k,-k'}
\end{equation}
Thus, gluing two open legs is implemented by contracting the corresponding
magnetic indices with the invariant tensor $T_{kk'}$.

\section{Entanglement Entropy of Superposed Link Spins}\label{app:superposed}

In this appendix we derive the formula
\begin{equation}
S_{l}
\;=\;
\sum_{i=1}^{i_{\mathrm{max}}} |a_i|^2 \,\log\bigl(2\,j(i)+1\bigr)
\;-\;\sum_{i=1}^{i_{\mathrm{max}}} |a_i|^2 \,\log\bigl(|a_i|^2\bigr)
\end{equation}
for the entanglement entropy of a single link $l$.  Recall that the most general normalized state on that link is
\begin{equation}
\lvert \varphi\rangle
\;=\;
\sum_{\,i=1}^{i_{\mathrm{max}}} a_i \,\lvert \varphi_{\,j(i)}\rangle,
\qquad
\sum_{i=1}^{i_{\mathrm{max}}} |a_i|^2 \;=\; 1
\end{equation}
where each basis vector \(\lvert \varphi_{\,j(i)}\rangle\) is given by
\begin{equation}
\lvert \varphi_{\,j(i)}\rangle
\;=\;
\frac{1}{\sqrt{\,2\,j(i)+1\,}}
\;\sum_{\,k=-\,j(i)}^{+\,j(i)}
\;\lvert k\rangle_{n_1}\,\lvert k\rangle_{n_2}
\end{equation}
Because each \(\lvert \varphi_{\,j(i)}\rangle\) is already in Schmidt form with rank \(2\,j(i)+1\), the overall superposition has Schmidt weights \(\{|a_i|^2\}\) on those sectors.  We now compute the reduced density matrix on node \(n_1\) and its von Neumann entropy.

\subsection{Reduced density matrix on node $n_1$}

The pure‐state density on the two‐node Hilbert space \(H_{n_1}\otimes H_{n_2}\) is
\begin{equation}
\lvert \varphi\rangle\langle \varphi\rvert
\;=\;
\sum_{i=1}^{i_{\mathrm{max}}}\sum_{\,i'=1}^{i_{\mathrm{max}}}
\;a_i\,a_{i'}^*
\;\lvert \varphi_{\,j(i)}\rangle\langle \varphi_{\,j(i')}\rvert
\end{equation}
We trace out the second factor (node \(n_2\)) to obtain
\begin{equation}
\rho_{\,n_1}
\;=\;
\mathrm{Tr}_{\,n_2}\bigl[\lvert \varphi\rangle\langle \varphi\rvert\bigr]
\;=\;
\sum_{i=1}^{i_{\mathrm{max}}}\sum_{\,i'=1}^{i_{\mathrm{max}}} a_i\,a_{i'}^*
\;\mathrm{Tr}_{\,n_2}\bigl[\lvert \varphi_{\,j(i)}\rangle\langle \varphi_{\,j(i')}\rvert\bigr]
\end{equation}
For fixed \(i,i'\), write
\begin{equation}
\lvert \varphi_{\,j(i)}\rangle\langle \varphi_{\,j(i')}\rvert
\;=\;
\frac{1}{\sqrt{(2\,j(i)+1)\,(2\,j(i')+1)}}\;
\sum_{\,m=-\,j(i)}^{+\,j(i)}
\sum_{\,n=-\,j(i')}^{+\,j(i')}
\;\Bigl(\lvert m\rangle_{n_1}\,\lvert m\rangle_{n_2}\Bigr)\,
\Bigl(\langle n\rvert_{n_1}\,\langle n\rvert_{n_2}\Bigr)
\end{equation}
Taking the partial trace over \(n_2\) kills off any terms with \(m \neq n\) or \(j(i)\neq j(i')\).  Hence
\begin{equation}
\mathrm{Tr}_{\,n_2}\bigl[\lvert \varphi_{\,j(i)}\rangle\langle \varphi_{\,j(i')}\rvert\bigr]
\;=\;
\delta_{\,j(i)\,,\,j(i')}\;
\frac{1}{\,2\,j(i)+1\,}\;
\sum_{\,m=-\,j(i)}^{+\,j(i)}
\;\lvert m\rangle_{n_1}\langle m\rvert_{\,n_1}
\end{equation}
Substituting back into \(\rho_{\,n_1}\) gives
\begin{equation}
\rho_{\,n_1}
\;=\;
\sum_{\,i=1}^{i_{\mathrm{max}}} |a_i|^2 
\;\frac{1}{\,2\,j(i)+1\,}\;
\sum_{\,m=-\,j(i)}^{+\,j(i)}
\;\lvert m\rangle_{n_1}\langle m\rvert_{\,n_1}
\end{equation}
Thus \(\rho_{\,n_1}\) has eigenvalues
\begin{equation}
\lambda_{\,i}
\;=\;
\frac{|a_i|^2}{\,2\,j(i)+1\,},
\quad
\text{each with multiplicity }2\,j(i)+1
\end{equation}
and we check normalization:
\begin{equation}
\sum_{\,i=1}^{i_{\mathrm{max}}} (2\,j(i)+1)\,\lambda_{\,i}
\;=\;
\sum_{\,i=1}^{i_{\mathrm{max}}} |a_i|^2
\;=\;1
\end{equation}

\subsection{Computation of $S(\rho_{n_1})$}

The von Neumann entropy is
\begin{equation}
S_{\ell}
\;=\;
-\,\mathrm{Tr}\bigl(\rho_{\,n_1}\,\log \rho_{\,n_1}\bigr)
\end{equation}
Since \(\rho_{\,n_1}\) has eigenvalues \(\lambda_i\) each repeated \(\,2\,j(i)+1\) times, its entropy is
\begin{equation}
S_{\ell}
\;=\;
\sum_{\,i=1}^{i_{\mathrm{max}}} (2\,j(i)+1)\,\bigl[\,-\,\lambda_{\,i}\,\log \lambda_{\,i}\bigr]
\end{equation}
Substitute \(\lambda_{\,i} = |a_i|^2/(2\,j(i)+1)\):
\begin{equation}
S_{\ell}
\;=\;
\sum_{\,i=1}^{i_{\mathrm{max}}} (2\,j(i)+1)\,
\Bigl[-\,\frac{|a_i|^2}{\,2\,j(i)+1\,}\;
\log\!\Bigl(\frac{|a_i|^2}{\,2\,j(i)+1\,}\Bigr)\Bigr]
\end{equation}
Thus,
\begin{align}
S_{\ell}
&=
\sum_{\,i=1}^{i_{\mathrm{max}}} (2\,j(i)+1)\,
\Bigl[-\,\frac{|a_i|^2}{\,2\,j(i)+1\,}\,\bigl(\log|a_i|^2 - \log(2\,j(i)+1)\bigr)\Bigr]
\nonumber\\
&=
\sum_{\,i=1}^{i_{\mathrm{max}}}
\bigl[\,-\,|a_i|^2\,\log|a_i|^2 \;+\; |a_i|^2\,\log(2\,j(i)+1)\bigr]
\end{align}
Reordering terms yields
\begin{equation}
S_{\ell}
\;=\;
\sum_{\,i=1}^{i_{\mathrm{max}}} |a_i|^2 \,\log\bigl(2\,j(i)+1\bigr)
\;-\;\sum_{\,i=1}^{i_{\mathrm{max}}} |a_i|^2 \,\log\bigl(|a_i|^2\bigr),
\end{equation}
which confirms Eq.~\eqref{eqn:superposed_entropy}.

\section{Calculation of Ising action}
\label{app:ising}

\subsection{Full Bulk Hilbert Space}
In this section we present details on the calculation of Eq.~\eqref{factored_action}. First, consider the term 
\begin{equation}
    \text{Tr}_L \big{[}\rho_L^{\otimes2} \bigotimes_{s_n=-1}\bigotimes_{i=1}^{V-1}S_n^i\big{]} = \text{Tr}_L \big{[}\big{(}\bigotimes_{l\in L}\big{(}\frac{1}{E_l}\sum_{a_l=0}^{E_l-1}\ket{a_l}\bra{a_l} \big{)}\big{)}^{\otimes2} \bigotimes_{s_n=-1}\bigotimes_{i=1}^{V-1}S_n^i\big{]}
\end{equation}
Let $m$ and $n$ be the endpoints of a link $l$. There are three distinct types of links depending on the values of the Ising spins $s_n$ and $s_m$. For calculational purpose, it is most convenient to treat the state $\ket{a_l}$ in its information-theoretical construction: $a_l$ maximally entangled pairs of qudits and $E_l-1-a_l$ nonentangled pairs of qudits. Thus we will use $ \ket{a_l}=\ket{\chi_d}^{\otimes a_l} \otimes \ket{00}^{\otimes(E_l-1-a_l)} $ where $\ket{\chi_d} = \frac{1}{\sqrt{d}}\sum_{\mu=0}^{d-1} \ket{\mu}_m\ket{\mu}_n$ is a pair of maximally entangled qudits.
\begin{enumerate}[label=(\arabic*)] 
  \item \textbf{No swap operator ($s_m=s_n=1$).}

    \begin{align}
    \begin{split}
    \mathrm{Tr_L}\bigl[\rho_l \otimes\rho_l] &= \frac{1}{E_l^2} \sum_{a=0}^{E_l-1}\sum_{a'=0}^{E_l-1}\mathrm{Tr_L}\bigl[\ket{a_l}\bra{a_l} \otimes \ket{a'_l}\bra{a'_l}] \\
    &= \frac{1}{E_l^2}\sum_{a=0}^{E_l-1}\sum_{a'=0}^{E_l-1} \sum_{\alpha=0}^{E_l-1}\sum_{\alpha'=0}^{E_l-1} \big{(}\bra{\alpha_l}\otimes\bra{\alpha'_l}\big{)}\ket{a_l}\bra{a_l} \otimes \ket{a'_l}\bra{a'_l}\big{(}\ket{\alpha_l}\otimes\ket{\alpha'_l}\big{)}\\
    &= \frac{1}{E_l^2}\sum_{a=0}^{E_l-1}\sum_{a'=0}^{E_l-1} \sum_{\alpha=0}^{E_l-1}\sum_{\alpha'=0}^{E_l-1} \big{(}\bra{\alpha_l}\otimes\bra{\alpha'_l}\big{)}\ip{a_l}{\alpha_l}\ip{a'_l}{\alpha'_l}\ket{a_l} \otimes \ket{a'_l} \\
    &=\frac{1}{E_l^2}\sum_{a=0}^{E_l-1}\sum_{a'=0}^{E_l-1} \sum_{\alpha=0}^{E_l-1}\sum_{\alpha'=0}^{E_l-1} \big{(}\bra{\alpha_l}\otimes\bra{\alpha'_l}\big{)}\delta_{a \alpha}\delta_{a' \alpha'} \ket{a_l}\otimes \ket{a'_l}\\
    &=\frac{1}{E_l^2}\sum_{a=0}^{E_l-1}\sum_{a'=0}^{E_l-1}\big{(}\bra{a_l}\otimes\bra{a'_l}\big{)}\ket{a_l} \otimes \ket{a'_l}\\
    &=1
    \end{split}
    \end{align}

  \item \textbf{One swap operator} ($s_m=-s_n$).

 \begin{align}
 \begin{split}
   \mathrm{Tr_L}\bigl[\rho_l \otimes\rho_lS_m] &= \frac{1}{E_l^2} \sum_{a=0}^{E_l-1}\sum_{a'=0}^{E_l-1}\mathrm{Tr_L}\bigl[\ket{a_l}\bra{a_l} \otimes \ket{a'_l}\bra{a'_l}S_m] \\
    &= \frac{1}{E_l^2}\sum_{a=0}^{E_l-1}\sum_{a'=0}^{E_l-1} \sum_{\alpha=0}^{E_l-1}\sum_{\alpha'=0}^{E_l-1} \big{(}\bra{\alpha_l}\otimes\bra{\alpha'_l}\big{)}\ket{a_l}\bra{a_l} \otimes \ket{a'_l}\bra{a'_l}S_m\big{(}\ket{\alpha_l}\otimes\ket{\alpha'_l}\big{)}\\
    &= \frac{1}{E_l^2}\sum_{a=0}^{E_l-1}\sum_{a'=0}^{E_l-1} \sum_{\alpha=0}^{E_l-1}\sum_{\alpha'=0}^{E_l-1} \big{(}\bra{\alpha_l}\otimes\bra{\alpha'_l}\big{)}\ip{\alpha_l}{a_l}\ip{\alpha'_l}{a'_l}S_m\big{(}\ket{a_l} \otimes \ket{a'_l}\big{)} \\
    &=\frac{1}{E_l^2}\sum_{a=0}^{E_l-1}\sum_{a'=0}^{E_l-1} \big{(}\bra{a_l}\otimes\bra{a'_l}\big{)}S_m\big{(}\ket{a_l} \otimes \ket{a'_l}\big{)}\\
    \end{split}
    \end{align}
    To calculate this expression, rewrite $\ket{a}$ as $\ket{a} = \bigotimes_{r=1}^{E_l-1}\ket{\eta_r^{(a)}}$ where
    \[
\ket{\eta_r^{(a)}}
=
\begin{cases}
\ket{\chi_d}, & r\le a,\\
\ket{00}, & r>a.
\end{cases}
\]
Factoring $S_m = \bigotimes_{r=1}^{E_l-1}S_{m,r}$ as well, we get
\begin{equation}  
(\bra{a}\otimes\bra{a'})S_m(\ket{a}\otimes \ket{a'})
=
\prod_{r=1}^{E_l-1}
\left[
\bra{\eta_r^{(a)}}\otimes\bra{\eta_r^{(a')}}
S_{m,r}
\ket{\eta_r^{(a)}}\otimes\ket{\eta_r^{(a')}}
\right]
\end{equation}
There are three cases to consider:
\begin{align}
 \begin{split}
(\bra{\chi_d}\otimes\bra{\chi_d})S_{m,r}(\ket{\chi_d}\otimes\ket{\chi_d})
&= d^{-2}\sum_{\mu_i,\mu_j,\mu_p,\mu_q=0}^{d-1}\bra{\mu_i}_m\bra{\mu_i}_n\bra{\mu_j}_m\bra{\mu_j}_nS_{m,r}\ket{\mu_p}_m\ket{\mu_p}_n\ket{\mu_q}_m\ket{\mu_q}_n\\
&=d^{-2}\sum_{\mu_i,\mu_j,\mu_p,\mu_q=0}^{d-1}\bra{\mu_i}_m\bra{\mu_i}_n\bra{\mu_j}_m\bra{\mu_j}_n\ket{\mu_q}_m\ket{\mu_p}_n\ket{\mu_p}_m\ket{\mu_q}_n\\
&=d^{-2}\sum_{\mu_i,\mu_j,\mu_p,\mu_q=0}^{d-1} \delta_{\mu_i,\mu_q}\delta_{\mu_i,\mu_p}\delta_{\mu_j,\mu_p}\delta_{\mu_j,\mu_q}\\
&=d^{-2}\sum_{\mu_i,\mu_j=0}^{d-1} \delta_{\mu_i,\mu_j}\delta_{\mu_j,\mu_i}\\
&=d^{-1},
    \end{split}
    \end{align}

\begin{align}
 \begin{split}
(\bra{\chi_d}\otimes\bra{00})S_{m,r}(\ket{\chi_d}\otimes\ket{00})
&= d^{-1}\sum_{\mu_i,\mu_p=0}^{d-1}\bra{\mu_i}_m\bra{\mu_i}_n\bra{0}_m\bra{0}_nS_{m,r}\ket{\mu_p}_m\ket{\mu_p}_n\ket{0}_m\ket{0}_n\\
&=d^{-1}\sum_{\mu_i,\mu_p=0}^{d-1}\bra{\mu_i}_m\bra{\mu_i}_n\bra{0}_m\bra{0}_n\ket{0}_m\ket{\mu_p}_n\ket{\mu_p}_m\ket{0}_n\\
&=d^{-1}\sum_{\mu_i,\mu_p=0}^{d-1}\delta_{\mu_i,0}\delta_{\mu_i,\mu_p}\delta_{0,\mu_p}\delta_{0,0}\\
&=d^{-1}\sum_{\mu_p=0}^{d-1}\delta_{0,\mu_p}\delta_{0,\mu_p}\\
&=d^{-1},
    \end{split}
    \end{align}
    and
\begin{equation}
    (\bra{00}\otimes\bra{00})S_{m,r}(\ket{00}\otimes\ket{00})=1.
\end{equation}
Therefore, if $r \le \max(a,a')$ we get a contribution of $d^{-1}$ and otherwise we get 1, which means
\begin{equation}
    (\bra{a}\otimes\bra{a'})S_m(\ket{a}\otimes \ket{a'}) = d^{-\max(a,a')}
\end{equation}
Substituting into the trace, we get

\begin{align}
 \begin{split}
   \mathrm{Tr_L}\bigl[\rho_l \otimes\rho_lS_m] &= \frac{1}{E_l^2} \sum_{a=0}^{E_l-1}\sum_{a'=0}^{E_l-1}d^{-\max(a,a')}\\
   &=\frac{1}{E_l^2} \sum_{a=0}^{E_l-1}(2a+1)d^{-a}
    \end{split}
    \end{align}

    By the form of the trace, we can interpret $\mathrm{Tr_L}\bigl[\rho_l \otimes\rho_lS_m]$ as computing the second Rényi entropy of the site $n$ in the total system described by $\rho_l$. We give it the name $e^{-s(l)}$. When $J_{min} \to \infty$ and hence $E_l \to \infty$, $s(l) \approx 2\log{E_l}$.

  \item \textbf{Two swap operators} ($s_m=s_n=-1$).

        \begin{align}
    \begin{split}
   \mathrm{Tr_L}\bigl[\rho_l \otimes\rho_lS_mS_n] &= \frac{1}{E_l^2} \sum_{a=0}^{E_l-1}\sum_{a'=0}^{E_l-1}\mathrm{Tr_L}\bigl[\ket{a_l}\bra{a_l} \otimes \ket{a'_l}\bra{a'_l}S_mS_n] \\
    &= \frac{1}{E_l^2}\sum_{a=0}^{E_l-1}\sum_{a'=0}^{E_l-1} \sum_{\alpha=0}^{E_l-1}\sum_{\alpha'=0}^{E_l-1} \big{(}\bra{\alpha_l}\otimes\bra{\alpha'_l}\big{)}\ket{a_l}\bra{a_l} \otimes \ket{a'_l}\bra{a'_l}S_mS_n\big{(}\ket{\alpha_l}\otimes\ket{\alpha'_l}\big{)}\\
    &= \frac{1}{E_l^2}\sum_{a=0}^{E_l-1}\sum_{a'=0}^{E_l-1} \sum_{\alpha=0}^{E_l-1}\sum_{\alpha'=0}^{E_l-1} \big{(}\bra{\alpha_l}\otimes\bra{\alpha'_l}\big{)}\ip{\alpha_l}{a_l}\ip{\alpha'_l}{a'_l}S_mS_n\big{(}\ket{a_l} \otimes \ket{a'_l}\big{)} \\
    &=\frac{1}{E_l^2}\sum_{a=0}^{E_l-1}\sum_{a'=0}^{E_l-1} \big{(}\bra{a_l}\otimes\bra{a'_l}\big{)}S_mS_n\big{(}\ket{a_l} \otimes \ket{a'_l}\big{)}\\
    &=\frac{1}{E_l^2}\sum_{a=0}^{E_l-1}\sum_{a'=0}^{E_l-1} \big{(}\bra{a_l}\otimes\bra{a'_l}\big{)}\big{(}\ket{a'_l} \otimes \ket{a_l}\big{)}\\
    &=\frac{1}{E_l^2}\sum_{a=0}^{E_l-1}\sum_{a'=0}^{E_l-1} \delta_{a,a'}^2\\
    &=\frac{1}{E_l}
    \end{split}
    \end{align}

\end{enumerate}

Multiplying the contribution from all links gives us the total contribution from the trace over $L$:
\begin{equation}
    \mathcal{A}(\{s_n\})_L = -\sum_{l\in L}\frac{s(l)}{2}(s_ms_n-1) + \sum_{l\in L}\frac{\log E_l}{4}(1-s_m)(1-s_n)
\end{equation}
Next we consider the term

\begin{equation}
    \text{Tr}_B\big{[}\rho_\partial^{\otimes2}S_B \bigotimes_{s_n=-1}S_n^\partial\big{]} = \text{Tr}_B\big{[} \bigotimes_{b\in B}\ket{l_b}\bra{l_b} \otimes \ket{l_b}\bra{l_b}S_B \bigotimes_{s_n=-1}S_n^\partial\big{]}
\end{equation}
Recall that the state $\ket{l_b}$ is a pair of maximally entangled $D$-dimensional qudits, which we write as $\ket{l_b}=\frac{1}{\sqrt{D}} \sum_{i=0}^{D-1}\ket{i}_{n_b}\ket{i}_b$. The basis vectors of this Hilbert space are $\ket{i}_{n_b}\ket{j}_b$. There are two cases to consider: $s_{n_b}=1$ and $s_{n_b}=-1$.

\begin{enumerate}[label=(\arabic*)] 
  \item $s_{n_b}=+1$
  \begin{align}
      \begin{split}
          \mathrm{Tr_L}\bigl[\ket{l_b}\bra{l_b}\otimes\ket{l_b}\bra{l_b}S_B] &= \sum_{ijkl}\bra{i}_{n_b}\bra{j}_{b}\bra{k}_{n_b}\bra{l}_{b}(\ket{l_b}\bra{l_b}\otimes\ket{l_b}\bra{l_b})S_B \ket{i}_{n_b}\ket{j}_{b}\ket{k}_{n_b}\ket{l}_{b}\\
          &=\frac{1}{D^2}\sum_{ijkl} \sum_{pqrs}\bra{i}_{n_b}\bra{j}_{b}\bra{k}_{n_b}\bra{l}_{b}\\&\;\;\;\;\;\;(\ket{p}_{n_b}\ket{p}_{b}\bra{q}_{n_b}\bra{q}_{l}\otimes \ket{r}_{n_b}\ket{r}_{b}\bra{s}_{n_b}\bra{s}_{l})\ket{i}_{n_b}\ket{l}_{b}\ket{k}_{n_b}\ket{j}_{b}\\
          &= \frac{1}{D^2}\sum_{ijkl} \sum_{pqrs}\bra{i}_{n_b}\bra{j}_{b}\bra{k}_{n_b}\bra{l}_{b} (\delta_{qi}\delta_{ql}\delta_{sk}\delta_{sj})\ket{p}_{n_b}\ket{p}_{b}\ket{r}_{n_b}\ket{r}_{b}\\
          &=\frac{1}{D^2}\sum_{ijkl} \sum_{pr}\bra{i}_{n_b}\bra{j}_{b}\bra{k}_{n_b}\bra{l}_{b} (\delta_{il}\delta_{jk})\ket{p}_{n_b}\ket{p}_{b}\ket{r}_{n_b}\ket{r}_{b}\\
          &=\frac{1}{D^2}\sum_{ijpr} \bra{i}_{n_b}\bra{j}_{b}\bra{j}_{n_b}\bra{i}_{b} \ket{p}_{n_b}\ket{p}_{b}\ket{r}_{n_b}\ket{r}_{b}\\
          &=\frac{1}{D^2}\sum_{ijpr} \delta_{ip}\delta_{jp}\delta_{jr}\delta_{ir}\\
          &= \frac{1}{D^2}\sum_{ij} \delta_{ij}\delta_{ij}\\
          &= \frac{1}{D^2}\sum_{i} 1\\
          &= \frac{1}{D}
      \end{split}
  \end{align}

\item $s_{n_b}=-1$
  \begin{align}
      \begin{split}
          \mathrm{Tr_L}\bigl[\ket{l_b}\bra{l_b}\otimes\ket{l_b}\bra{l_b}S_{n_b}S_B] &= \sum_{ijkl}\bra{i}_{n_b}\bra{j}_{b}\bra{k}_{n_b}\bra{l}_{b}(\ket{l_b}\bra{l_b}\otimes\ket{l_b}\bra{l_b})S_{n_b}S_B \ket{i}_{n_b}\ket{j}_{b}\ket{k}_{n_b}\ket{l}_{b}\\
          &=\frac{1}{D^2}\sum_{ijkl} \sum_{pqrs}\bra{i}_{n_b}\bra{j}_{b}\bra{k}_{n_b}\bra{l}_{b}\\&\;\;\;\;\;\;(\ket{p}_{n_b}\ket{p}_{b}\bra{q}_{n_b}\bra{q}_{l}\otimes \ket{r}_{n_b}\ket{r}_{b}\bra{s}_{n_b}\bra{s}_{l})\ket{k}_{n_b}\ket{l}_{b}\ket{i}_{n_b}\ket{j}_{b}\\
          &= \frac{1}{D^2}\sum_{ijkl} \sum_{pqrs}\bra{i}_{n_b}\bra{j}_{b}\bra{k}_{n_b}\bra{l}_{b} (\delta_{qk}\delta_{ql}\delta_{si}\delta_{sj})\ket{p}_{n_b}\ket{p}_{b}\ket{r}_{n_b}\ket{r}_{b}\\
          &=\frac{1}{D^2}\sum_{ijkl} \sum_{pr}\bra{i}_{n_b}\bra{j}_{b}\bra{k}_{n_b}\bra{l}_{b} (\delta_{ij}\delta_{kl})\ket{p}_{n_b}\ket{p}_{b}\ket{r}_{n_b}\ket{r}_{b}\\
          &=\frac{1}{D^2}\sum_{ijpr} \bra{i}_{n_b}\bra{i}_{b}\bra{k}_{n_b}\bra{k}_{b} \ket{p}_{n_b}\ket{p}_{b}\ket{r}_{n_b}\ket{r}_{b}\\
          &=\frac{1}{D^2}\sum_{ijpr} \delta_{ip}\delta_{ip}\delta_{kr}\delta_{kr}\\
          &= \frac{1}{D^2}\sum_{ik} 1\\
          &= 1
      \end{split}
  \end{align}

\end{enumerate}
Combining the contributions, we get

\begin{equation}
    \mathcal{A}(\{s_n\})_\mathscr{\partial} = \frac{1}{2}\sum_{b\in B} (s_{n_b}+1) \log{D}
\end{equation}

Finally we consider the term
\begin{equation}
    \text{Tr}_{\mathscr{I}}\big{[}\bigotimes_{s_n=-1} S_n^0\big{]}
\end{equation}
When $s_n=1$ this just becomes $\text{Tr}_{\mathscr{I}}\big{[}\mathbbm{1} \otimes \mathbbm{1}\big{]} $, which is square of the dimension of the intertwiner space, $D_{\mathscr{I}_n}^2$. When $s_n=-1$, only basis states in the form of an intertwiner state tensored with itself gives $1$, so the trace is $ D_{\mathscr{I}_n}$. Hence the contribution is 
\begin{equation}
    \mathcal{A}(\{s_n\})_\mathscr{I} = -\sum_{n\in N}\frac{1}{2}(s_n+3)\log{D_{\mathscr{I}_n}}
\end{equation}
Thus the total action of the effective Ising model is
\begin{align}
\begin{split}
     \mathcal{A}(\{s_n\}) &= -\sum_{l\in L}\frac{s(l)}{2}(s_ms_n-1)+ \sum_{l\in L}\frac{\log E_l}{4}(1-s_m)(1-s_n)\\&+\frac{1}{2}\sum_{b\in B} (s_{n_b}+1) \log{D} -\sum_{n\in N}\frac{1}{2}(s_n+3)\log{D_{\mathscr{I}_n}}
\end{split}
\end{align}
We can shift it by the constant $2\sum_{n\in N} \log D_{\mathscr{I}_n}$ to obtain
\begin{align}
\begin{split}
     \mathcal{A}_1(\{s_n\}) &= -\sum_{l\in L}\frac{s(l)}{2}(s_ms_n-1) + \sum_{l\in L}\frac{\log E_l}{4}(1-s_m)(1-s_n)\\&+\frac{1}{2}\sum_{b\in B} (s_{n_b}-1) \log{D} -\sum_{n\in N}\frac{1}{2}(s_n-1)\log{D_{\mathscr{I}_n}}+|B|\log{D}
\end{split}
\end{align}

\subsection{Semiclassical Subspace}

The only term that's different from the calculation in the previous section is 
\begin{equation}
    \text{Tr}_L \big{[}\rho_L^{\otimes2} \bigotimes_{s_n=-1}\bigotimes_{i=1}^{V-1}S_n^i\big{]}
\end{equation}
We again have three cases to consider. 

\begin{enumerate}[label=(\arabic*)] 
  \item \textbf{No swap operator ($s_m=s_n=1$).}

For links with classical entanglement $a_{nm}=0$:

    \begin{align}
    \begin{split}
    \mathrm{Tr_L}\bigl[\rho_l \otimes\rho_l] &= \frac{1}{(2\Lambda+1)^2} \sum_{a=0}^{2\Lambda}\sum_{a'=0}^{2\Lambda}\mathrm{Tr_L}\bigl[\ket{a_l}\bra{a_l} \otimes \ket{a'_l}\bra{a'_l}] \\
    &= \frac{1}{(2\Lambda+1)^2}\sum_{a=0}^{2\Lambda}\sum_{a'=0}^{2\Lambda} \sum_{\alpha=0}^{2\Lambda}\sum_{\alpha'=0}^{2\Lambda} \big{(}\bra{\alpha_l}\otimes\bra{\alpha'_l}\big{)}\ket{a_l}\bra{a_l} \otimes \ket{a'_l}\bra{a'_l}\big{(}\ket{\alpha_l}\otimes\ket{\alpha'_l}\big{)}\\
    &= \frac{1}{(2\Lambda+1)^2}\sum_{a=0}^{2\Lambda}\sum_{a'=0}^{2\Lambda} \sum_{\alpha=0}^{2\Lambda}\sum_{\alpha'=0}^{2\Lambda} \big{(}\bra{\alpha_l}\otimes\bra{\alpha'_l}\big{)}\ip{a_l}{\alpha_l}\ip{a'_l}{\alpha'_l}\ket{a_l} \otimes \ket{a'_l} \\
    &=\frac{1}{(2\Lambda+1)^2}\sum_{a=0}^{2\Lambda}\sum_{a'=0}^{2\Lambda} \sum_{\alpha=0}^{2\Lambda}\sum_{\alpha'=0}^{2\Lambda} \big{(}\bra{\alpha_l}\otimes\bra{\alpha'_l}\big{)}\delta_{a \alpha}\delta_{a' \alpha'} \ket{a_l}\otimes \ket{a'_l}\\
    &=\frac{1}{(2\Lambda+1)^2}\sum_{a=0}^{2\Lambda}\sum_{a'=0}^{2\Lambda}\big{(}\bra{a_l}\otimes\bra{a'_l}\big{)}\ket{a_l} \otimes \ket{a'_l}\\
    &=1
    \end{split}
    \end{align}

For links with classical entanglement $a_{nm}\not=0$, just replace the bounds of the sum by $a_{nm}-\Lambda$ to $a_{nm}+\Lambda$, and the result is still $1$.
    
  \item \textbf{One swap operator} ($s_m=-s_n$).

  For links with classical entanglement $a_{nm}=0$:

    \begin{align}
    \begin{split}
   \mathrm{Tr_L}\bigl[\rho_l \otimes\rho_lS_m] &= \frac{1}{(2\Lambda+1)^2} \sum_{a=0}^{2\Lambda}\sum_{a'=0}^{2\Lambda}\mathrm{Tr_L}\bigl[\ket{a_l}\bra{a_l} \otimes \ket{a'_l}\bra{a'_l}S_m] \\
    &= \frac{1}{(2\Lambda+1)^2}\sum_{a=0}^{2\Lambda}\sum_{a'=0}^{2\Lambda} \sum_{\alpha=0}^{2\Lambda}\sum_{\alpha'=0}^{2\Lambda} \big{(}\bra{\alpha_l}\otimes\bra{\alpha'_l}\big{)}\ket{a_l}\bra{a_l} \otimes \ket{a'_l}\bra{a'_l}S_m\big{(}\ket{\alpha_l}\otimes\ket{\alpha'_l}\big{)}\\
    &= \frac{1}{(2\Lambda+1)^2}\sum_{a=0}^{2\Lambda}\sum_{a'=0}^{2\Lambda} \sum_{\alpha=0}^{2\Lambda}\sum_{\alpha'=0}^{2\Lambda} \big{(}\bra{\alpha_l}\otimes\bra{\alpha'_l}\big{)}\ip{\alpha_l}{a_l}\ip{\alpha'_l}{a'_l}S_m\big{(}\ket{a_l} \otimes \ket{a'_l}\big{)} \\
    &=\frac{1}{(2\Lambda+1)^2}\sum_{a=0}^{2\Lambda}\sum_{a'=0}^{2\Lambda} \big{(}\bra{a_l}\otimes\bra{a'_l}\big{)}S_m\big{(}\ket{a_l} \otimes \ket{a'_l}\big{)}\\
    &=\frac{1}{(2\Lambda+1)^2}\sum_{a=0}^{2\Lambda}\sum_{a'=0}^{2\Lambda} d^{-\max(a,a')}\\
    &=\frac{1}{(2\Lambda+1)^2} \sum_{a=0}^{2\Lambda}(2a+1)d^{-a}
    \end{split}
    \end{align}

    Similar to the previous section, we denote this result by $e^{-s_0(l)}$. If $\Lambda \gg 1$, $s_0(l) \approx 2\log{(2\Lambda+1)}$.

    For links with classical entanglement $a_{nm}\not=0$, just replace the bounds of the sum by $a_{nm}-\Lambda$ to $a_{nm}+\Lambda$, and the result is

    \begin{equation}
        \mathrm{Tr_L}\bigl[\rho_l \otimes\rho_lS_m] = \frac{1}{(2\Lambda+1)^2}\sum_{a=a_{nm}-\Lambda}^{a_{nm}+\Lambda}(2(a-a_{nm}+\Lambda)+1)d^{-a}=\frac{d^{-(a_{nm}-\Lambda)}}{(2\Lambda+1)^2} \sum_{a=0}^{2\Lambda}(2a+1)d^{-a}
    \end{equation}

    We denote this result by $e^{-s_1(l)}$. If $a_{nm}\gg \Lambda \gg 1$, $s_1(l) \approx (a_{nm}-\Lambda)\log d+2\log{(2\Lambda+1)}$.

  \item \textbf{Two swap operators} ($s_m=s_n=-1$).

For links with classical entanglement $a_{nm}=0$:

        \begin{align}
    \begin{split}
   \mathrm{Tr_L}\bigl[\rho_l \otimes\rho_lS_mS_n] &= \frac{1}{(2\Lambda+1)^2} \sum_{a=0}^{2\Lambda}\sum_{a'=0}^{2\Lambda}\mathrm{Tr_L}\bigl[\ket{a_l}\bra{a_l} \otimes \ket{a'_l}\bra{a'_l}S_mS_n] \\
    &= \frac{1}{(2\Lambda+1)^2}\sum_{a=0}^{2\Lambda}\sum_{a'=0}^{2\Lambda} \sum_{\alpha=0}^{2\Lambda}\sum_{\alpha'=0}^{2\Lambda} \big{(}\bra{\alpha_l}\otimes\bra{\alpha'_l}\big{)}\ket{a_l}\bra{a_l} \otimes \ket{a'_l}\bra{a'_l}S_mS_n\big{(}\ket{\alpha_l}\otimes\ket{\alpha'_l}\big{)}\\
    &= \frac{1}{(2\Lambda+1)^2}\sum_{a=0}^{2\Lambda}\sum_{a'=0}^{2\Lambda} \sum_{\alpha=0}^{2\Lambda}\sum_{\alpha'=0}^{2\Lambda} \big{(}\bra{\alpha_l}\otimes\bra{\alpha'_l}\big{)}\ip{\alpha_l}{a_l}\ip{\alpha'_l}{a'_l}S_mS_n\big{(}\ket{a_l} \otimes \ket{a'_l}\big{)} \\
    &=\frac{1}{(2\Lambda+1)^2}\sum_{a=0}^{2\Lambda}\sum_{a'=0}^{2\Lambda} \big{(}\bra{a_l}\otimes\bra{a'_l}\big{)}S_mS_n\big{(}\ket{a_l} \otimes \ket{a'_l}\big{)}\\
    &=\frac{1}{(2\Lambda+1)^2}\sum_{a=0}^{2\Lambda}\sum_{a'=0}^{2\Lambda} \big{(}\bra{a_l}\otimes\bra{a'_l}\big{)}\big{(}\ket{a'_l} \otimes \ket{a_l}\big{)}\\
    &=\frac{1}{(2\Lambda+1)^2}\sum_{a=0}^{2\Lambda}\sum_{a'=0}^{2\Lambda} \delta_{a,a'}^2\\
    &=\frac{1}{2\Lambda+1}
    \end{split}
    \end{align}

For links with classical entanglement $a_{nm}\not=0$, just replace the bounds of the sum by $a_{nm}-\Lambda$ to $a_{nm}+\Lambda$, and the result is still $\frac{1}{2\Lambda+1}$. 

\end{enumerate}

The total action of the effective Ising model is
\begin{align}
    \begin{split}
        \mathcal{A}(\{s_n\}) &= -\sum_{l\in L_0}\frac{s_0(l)}{2}(s_ms_n-1)-\sum_{l\in L_1}\frac{s_1(l)}{2}(s_ms_n-1)+ \sum_{l\in L}\frac{\log(2\Lambda+1)}{4}(1-s_m)(1-s_n) \\&+\frac{1}{2}\sum_{b\in B} (s_{n_b}+1) \log{D} -\sum_{n\in N}\frac{1}{2}(s_n+3)\log{D_{\mathscr{I}_n}}
    \end{split}
\end{align}
Shifting by the constant $2\sum_{n\in N} \log D_{\mathscr{I}_n}$, we obtain
\begin{align}
    \begin{split}
        \mathcal{A}(\{s_n\}) &= -\sum_{l\in L_0}\frac{s_0(l)}{2}(s_ms_n-1)-\sum_{l\in L_1}\frac{s_1(l)}{2}(s_ms_n-1)+\sum_{l\in L}\frac{\log(2\Lambda+1)}{4}(1-s_m)(1-s_n)\\ &+ \frac{1}{2}\sum_{b\in B} (s_{n_b}-1) \log{D} -\sum_{n\in N}\frac{1}{2}(s_n-1)\log{D_{\mathscr{I}_n}}+|B|\log{D}
    \end{split}
\end{align}

\section{Construction of Intertwiner Subspace}
\label{app:intertwiners}

In this appendix we give a detailed account of how the “small‐fluctuation’’ subspace of intertwiners is defined at each node, once the adjacent spins have been restricted to lie within their respective ranges.  Our goal is to exhibit a choice of basis for each $\mathscr{I}_{n}$ that diagonalizes all $(N-1)$ incident spin operators simultaneously, and to show that for each such assignment of spins there is exactly one allowed basis vector.  As a result, specifying all node‐intertwiners amounts precisely to specifying the array of spins on every adjacent link.  Throughout, we denote by
\begin{equation}
  \mathscr{I}_{n}\bigl(a_{n\bullet}\bigr)
  \;=\;
  \mathrm{Inv}_{SU(2)}\Bigl(\,\bigotimes_{m\neq n}V_{\,a_{nm}}\Bigr)
\end{equation}
the invariant subspace at node $n$ for a fixed choice of incident spins $\{a_{nm}\}_{m\neq n}$.  We now explain how to construct a basis of $\mathscr{I}_{n}$ in which every operator $\hat F_{\,n\to m}$ (measuring the spin on the leg between $n$ and $m$) is diagonal.

\subsection{Flux operators and simultaneous diagonalization}

Each leg $(n\leftrightarrow m)$ carries an $\mathrm{SU}(2)$ representation $V_{\,a_{nm}}$, and on that factor we have the usual angular‐momentum generators with $a_{mn}$ being treated as the effective spin variables
\begin{equation*}
  \bigl\{\hat J^x_{\,n\to m},\,\hat J^y_{\,n\to m},\,\hat J^z_{\,n\to m}\bigr\}
\end{equation*}
whose Casimir satisfies 
\begin{equation}
  \hat{\mathbf J}^2_{\,n\to m}\ket{a_{nm}} = a_{nm}(a_{nm}+1)\ket{a_{nm}}
\end{equation}
We define 
\begin{equation}
  \hat{\mathbf F}_{\,n\to m} \coloneqq \hat{\mathbf J}_{\,n\to m}, 
  \qquad
  \hat F_{\,n\to m} \coloneqq\sqrt{\hat{\mathbf F}^2_{\,n\to m}}
\end{equation}
so that any vector in $V_{\,a_{nm}}$ is an eigenvector of $\hat F_{\,n\to m}$ with eigenvalue $a_{nm}$.  Because the node’s intertwiner space $\mathscr{I}_{n}(a_{n\bullet})$ lies inside $\bigotimes_{m\neq n}V_{\,a_{nm}}$,
we have one flux‐magnitude operator $\hat F_{\,n\to m}$ for each of the $(N-1)$ incident legs.  Our construction requires that, for each choice of the $(N-1)$ spin‐values $a_{n1},\dots,a_{n,n-1},a_{n,n+1},\dots,a_{nN}$, there is a unique vector in $\mathscr{I}_{n}$ on which
\begin{equation}
  \hat F_{\,n\to m}\,\ket{\iota_{n}(a_{n\bullet})} 
  \;=\; a_{nm}\,\ket{\iota_{n}(a_{n\bullet})}
  \quad
  \text{for each }m\neq n
\end{equation}
Because the operators $\{\hat F_{\,n\to m}\}_{m\neq n}$ fail to commute as full vector operators, the exact definition of “simultaneous eigenvector” is implemented by working in the recoupling basis, as follows.

\subsection{Recoupling‐basis construction}

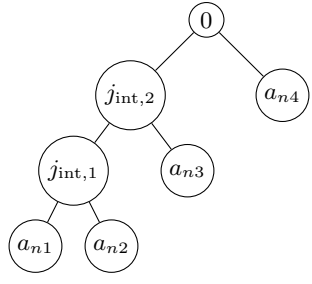
\begin{figure}
    \centering
    \begin{tikzpicture}[
    level distance=1cm,
    level 1/.style={sibling distance=2cm},
    level 2/.style={sibling distance=1.5cm},
    level 3/.style={sibling distance=1cm},
    every node/.style={font=\footnotesize, draw, circle, inner sep=2pt}
  ]
  \node (root) {0}
    child { node (j2) {$j_{\mathrm{int},2}$}
      child { node (j1) {$j_{\mathrm{int},1}$}
        child { node {$a_{n1}$} }
        child { node {$a_{n2}$} }
      }
      child { node {$a_{n3}$} }
    }
    child { node {$a_{n4}$} };
\end{tikzpicture}
    \caption{Recoupling tree for a 4‐valent node.  Incoming spins $a_{n1},a_{n2},a_{n3},a_{n4}$ are first coupled pairwise: $a_{n1}\otimes a_{n2}\to j_{\mathrm{int},1}$, then $j_{\mathrm{int},1}\otimes a_{n3}\to j_{\mathrm{int},2}$, and finally $j_{\mathrm{int},2}\otimes a_{n4}\to 0$.  Each intermediate label $j_{\mathrm{int},i}$ is uniquely fixed by requiring closure to total spin zero.}

    \label{fig:coupling_tree}
\end{figure}

To diagonalize all $\hat F_{\,n\to m}$ at once, we fix a binary coupling order among the $(N-1)$ legs at node $n$.  Concretely, one chooses an ordered pairing:
\begin{equation}
  \bigl((\,\cdots((\,a_{n1}\otimes a_{n2}\to j_{\rm int,1}\!)\otimes a_{n3}\to j_{\rm int,2}\!)\cdots)\otimes a_{n(N-1)}\to 0\bigr)
\end{equation}
so that at each step we couple two representations down to a new intermediate spin, until the final total is zero.  In this scheme:

\begin{itemize}
  \item \textbf{First coupling.}  Pair $a_{n1}\otimes a_{n2}\to j_{\rm int,1}$, with 
  \begin{equation}
    \bigl|\,a_{n1}-a_{n2}\bigr|\;\le\; j_{\rm int,1}\;\le\; a_{n1}+a_{n2}
  \end{equation}
  \item \textbf{Second coupling.}  Pair $j_{\rm int,1}\otimes a_{n3}\to j_{\rm int,2}$, with 
  \begin{equation}
    \bigl|\,j_{\rm int,1}-a_{n3}\bigr|\;\le\; j_{\rm int,2}\;\le\; j_{\rm int,1}+a_{n3}
  \end{equation}
  \item Continue in the same way, at each step coupling the previous intermediate spin with the next leg, until the final coupling 
  \begin{equation*}
    j_{\rm int,(N-2)} \;\otimes\; a_{n(N-1)} \;\longrightarrow\; 0
  \end{equation*}
  forces $j_{\rm int,(N-2)} = a_{n(N-1)}$. A coupling tree for a 4-valent node is shown in figure~\ref{fig:coupling_tree}.
\end{itemize}

Every choice of consistent intermediate spins 
\begin{equation*}
  \{\,j_{\rm int,1},\,j_{\rm int,2},\dots,j_{\rm int,(N-2)}\}
\end{equation*}
that satisfies the triangle‐inequalities at each step of the tree yields one orthonormal basis vector
\begin{equation}
  \ket{\,a_{n1},\,a_{n2},\,j_{\rm int,1},\,a_{n3},\,j_{\rm int,2},\,\dots,a_{n(N-1)}\!}
\end{equation}
in $\mathscr{I}_{n}(a_{n\bullet})$.  In particular, if we demand that the flux magnitudes on all $(N-1)$ legs equal exactly 
\(\{\,a_{n1},\,a_{n2},\dots,a_{n(N-1)}\}\), 
then the final coupling forces 
\begin{equation}
  j_{\rm int,(N-2)} = a_{n(N-1)}
\end{equation} 
In a general recoupling tree, the prior coupling spin $j_{\rm int,(N-3)}$ is not unique and satisfies
\begin{equation}
  0=\bigl|\,j_{\rm int,(N-2)} - a_{n(N-1)}\bigr| \le  j_{\rm int,(N-3)} \le j_{\rm int,(N-2)}+a_{n(N-1)} = 2a_{n(N-1)}
\end{equation}
However, here we require that the intertwiner matches the intrinsic data of a classical polyhedron, which is possible only for large spins. Since the face areas are fixed, the extrinsic geometry remains totally uncertain. The proper method to implement this condition is the Livine-Speziale (LS) coherent intertwiner~\cite{Livine:2007vk}. Under this prescription, the 
classical polyhedron is described by $\{(a_{mn},\mathbf n_{nm}^{(0)})\}$ where $\mathbf n_{nm}^{(0)}$ is the unit normal vector of each surface and $a_{mn}$ is the area of the corresponding surface. In order for them to form a polyhedron, they must satisfy the closure condition $\sum a_{nm}\,\mathbf n_{nm}^{(0)} \;=\;\mathbf0$~\cite{Rovelli:1995ac}. Intermediate coupling spins in the recoupling tree can only take extremal values (here $0$ or $2a_{n(N-1)}$ for $j_{\rm int,(N-3)}$ ), and which one to take depends on whether the unit normal $\mathbf n_{N-3}^{(0)}$ anti-aligns or aligns with the partial sum $\sum a_{nm}\,\mathbf n_{nm}^{(0)}$ up to $a_{n(N-4)}$. This prescription gives only one possible value for $j_{\rm int,(N-3)}$, and we can work backwards from there and
 select a distinguished semiclassical center in $\mathscr{I}_{n}(a_{n\bullet})$ in the end.  We denote that unique vector by
\begin{equation}
  \ket{\iota_{n}\bigl(a_{n\bullet}\bigr)} 
  \;\in\; 
  \mathscr{I}_{n}\bigl(a_{n\bullet}\bigr)
\end{equation}
and by construction it satisfies 
\begin{equation}
  \hat F_{\,n\to m}\,\ket{\iota_{n}(a_{n\bullet})}
  \;=\;
  a_{nm}\,\ket{\iota_{n}(a_{n\bullet})},
  \quad
  \sum_{m\neq n}a_{nm}\,\mathbf n_{nm}^{(0)} \;=\;\mathbf0
\end{equation}
In short, the single recoupling‐basis vector $\ket{\iota_{n}(a_{n\bullet})}$ is exactly the simultaneous eigenvector of all flux‐magnitudes $\hat F_{\,n\to m}$ with eigenvalues $a_{nm}$.

\subsection{Handling zero‐background legs}

When a given link $(n\leftrightarrow m)$ has $a_{nm}^{(0)}=0$ in the classical geometry, our prescription nonetheless allows $a_{nm}$ to fluctuate in $[0,\,2\Lambda]$.  In that case, $V_{\,a_{nm}}$ is a nontrivial $(2\,a_{nm}+1)$‐dimensional representation, and the recoupling basis still makes sense: one simply regards that leg as carrying spin $a_{nm}\le2\Lambda$.  The four originally nonzero legs at node $n$ continue to be of order $J\gg\Lambda$; the additional “formerly zero’’ legs each carry a small spin at most $2\Lambda$, which is always admissible when coupling to the dominant legs.  In particular:

\begin{enumerate}
  \item Exact closure in the recoupling‐basis sense continues to hold because one can always choose intermediate spins so that the final coupling yields total spin zero.  The only requirement is that at each step of the binary tree, the pair of spins obey the triangle‐inequalities; since the large spins are of order $J\gg2\Lambda$, and the small spins are no larger than $2\Lambda$, there is never a conflict.
  \item Although having zero‐background legs fluctuate up to $2\Lambda$ could in principle disturb a classical closure among the four large legs by an amount $\mathcal O(\Lambda)$, the recoupling basis automatically adjusts intermediate spins to restore exact closure.  Equivalently, one is choosing the unique invariant vector in $\mathscr{I}_{n}$ labeled by all $(N-1)$ spin‐values $\{\,a_{nm}\}$, regardless of whether those spins originally came from “large” or “small” background assignments.
\end{enumerate}

Thus, for every allowed $(N-1)$-tuple of incident labels, whether some of the corresponding legs vanish in the background geometry or not, the construction above determines a distinguished semiclassical intertwiner
\begin{equation}
\ket{\iota_n(a_{n\bullet})}
\in
\mathcal I_n(a_{n\bullet}),
\end{equation}
peaked on the polyhedron compatible with those data. This vector should be understood as the center of the corresponding semiclassical packet. Fixing the data carried by the incident links determines the classical face configuration, but it does not eliminate nearby quantum fluctuations of the polyhedral shape encoded by the intertwiner.

The link and intertwiner fluctuations have distinct physical meanings. The link variables specify the amount of entanglement assigned to each internal face and therefore control how strongly the two adjacent polyhedra are glued. The intertwiner sector instead describes how the faces meeting at a node are assembled into a polyhedral shape, including fluctuations of their relative orientations, recoupling channels, and volume. The closure condition correlates these shape fluctuations with the incident link data, but it does not identify the two sets of degrees of freedom.

Within the small-fluctuation regime, the admissibility conditions remain stable. Around each distinguished semiclassical intertwiner, we therefore retain a finite packet of compatible shape states. The corresponding shape-response label has $2\Lambda+1$ allowed values for each internal face. Since the two endpoints of an internal link describe the same shared face, compatibility across the link identifies their labels. The independent intertwiner fluctuations are consequently counted once per internal link, rather than once per endpoint. The dimension of the compatible intertwiner sector is therefore
\begin{equation}
D_{\mathscr I}=(2\Lambda+1)^{|L|}=(2\Lambda+1)^{\frac{1}{2}|N|(|N|-1)} 
\label{eq:intertwiner-subspace-dimension}
\end{equation}

Independently, the link-entanglement variables fluctuate within the same windows, giving $2\Lambda+1$ orthogonal link states for each internal link. The link sector therefore also has dimension $(2\Lambda+1)^{|L|}$. For every allowed link-entanglement configuration, the compatible intertwiner degrees of freedom span a $D_{\mathscr I}$-dimensional semiclassical packet. The full semiclassical subspace consequently has dimension
\begin{equation}
\dim \mathcal H({a_l})=
(2\Lambda+1)^{2|L|} =
(2\Lambda+1)^{|N|(|N|-1)} \
\label{eq:semiclassical-subspace-dimension}
\end{equation}

The square reflects two independent but mutually compatible components of the semiclassical data. One factor counts fluctuations in the amount of entanglement assigned to the internal faces, while the other counts fluctuations of the polyhedral shapes compatible with those face data and with closure at the nodes. This decomposition is precisely mirrored by the two contributions to the all-down Ising action in Eq. \ref{eq:A_1min}.

\section{Overlap between Classical States}\label{app:overlap}

In this section we present the calculation of $\avg{|C_{ab}|^2}$ used in Section~\ref{sec:overlap}. We begin with
\begin{equation}
    \avg{|C_{ab}|^2}=\text{Tr}_{L,\mathscr{I},B}[\bigotimes_{l\in L}\ket{a_l}\bra{b_l}\otimes \ket{b_l}\bra{a_l}\bigotimes_{n\in N}\avgn{\ket{n}\bra{n}\otimes \ket{n}\bra{n}}\bigotimes_{b\in B}\ket{l_b}\bra{l_b}\otimes\ket{l_b}\bra{l_b}]
\end{equation}
Similar to section~\ref{sec:Btb_iso}, we evaluate the average over $\ket{n}$ using Eq.~\eqref{vertex_avg} to get

\begin{align}
    \begin{split}
        \avg{|C_{ab}|^2}&=\bigg{(}\prod_{n\in N}\frac{1}{D_n(D_n+1)}\bigg{)}\text{Tr}_{L,\mathscr{I},B}[\bigotimes_{l\in L}\ket{a_l}\bra{b_l}\otimes \ket{b_l}\bra{a_l}\bigotimes_{n\in N}(\mathbbm{1}_n+S_n)\bigotimes_{b\in B}\ket{l_b}\bra{l_b}\otimes\ket{l_b}\bra{l_b}]\\
        &=\bigg{(}\prod_{n\in N}\frac{1}{D_n(D_n+1)}\bigg{)}\sum_{A\in2^{N}}\text{Tr}_{L,\mathscr{I},B}[S_A\bigotimes_{l\in L}\ket{a_l}\bra{b_l}\otimes \ket{b_l}\bra{a_l}\bigotimes_{b\in B}\ket{l_b}\bra{l_b}\otimes\ket{l_b}\bra{l_b}]
    \end{split}
\end{align}
Since the inside of the trace no longer depends on the intertwiner, the trace over $\mathscr{I}$ gives the dimension of the spin-network vertex space $\prod_{n\in N} D_n$ , which we absorb into a multiplicative prefactor $\tilde{C}$. Like what we did in Section~\ref{sec:Btb_iso}, we factor the swap operator $S_n$ into bulk links $S_n^i$ and boundary links $S_n^\partial$ and define $\rho_\partial = \bigotimes_{b\in B}\ket{l_b}\bra{l_b}$. After factoring the trace, we get
\begin{equation}
    \avg{|C_{ab}|^2}= \tilde{C}\sum_{A\in2^{N}}\text{Tr}_{L}[\bigotimes_{l\in L}\ket{a_l}\bra{b_l}\otimes \ket{b_l}\bra{a_l}S_A^i]\,\text{Tr}_{B}[\rho_B^{\otimes 2}\,S_A^\partial]
\end{equation}
where $S_A^i$ stands for $ \bigotimes_{n\in A}\bigotimes_{i=1}^{V-1}S_n^i$. The trace over $B$ is identical to a term appearing in Section~\ref{sec:Btb_iso} and has been computed in Appendix~\ref{app:ising}. The result is
\begin{equation}
    \text{Tr}_{B}[\rho_B^{\otimes 2}\,S_A^\partial] = D^{-|A \cap B|}
\end{equation}
where $|A \cap B|$ is the number of nodes in $A$ that are connected to a boundary node. Substituting this into the previous expression gives us
\begin{equation}
    \avg{|C_{ab}|^2}= \tilde{C}\sum_{A\in2^{N}}\text{Tr}_{L}[\bigotimes_{l\in L}\ket{a_l}\bra{b_l}\otimes \ket{b_l}\bra{a_l}S_A^i]\,D^{-|A \cap B|}
\end{equation}
Next, notice that $S_A^i = S_N^iS_{A^c}^i$ where $A^c = N -A$ is the complement of $A$. In addition, $|A \cap B| = |B| - |A^c\cap B|$. Thus we obtain
\begin{align}
    \begin{split}
        \avg{|C_{ab}|^2}&=\tilde{C}\sum_{A\in2^{N}}\text{Tr}_{L}[\bigotimes_{l\in L}\ket{a_l}\bra{b_l}\otimes \ket{b_l}\bra{a_l}S_N^iS_{A^c}^i]\,D^{ |A^c\cap B|-|B|}\\
        &=\tilde{C}\sum_{A\in2^{N}}\text{Tr}_{L}[\bigotimes_{l\in L}\ket{a_l}\bra{a_l}\otimes \ket{b_l}\bra{b_l}S_{A^c}^i]\,D^{ |A^c\cap B|-|B|}
    \end{split}
\end{align}
Note that summing over all $A$'s in the power set is the same as summing over its complement $A^c$. Thus we can write the sum over $A^c$, then relabel $A^c$ back to $A$ and arrive at
\begin{equation}
    \avg{|C_{ab}|^2}=\tilde{C}\sum_{A\in2^{N}}\text{Tr}_{L}[\bigotimes_{l\in L}\ket{a_l}\bra{a_l}\otimes \ket{b_l}\bra{b_l}S_{A}^i]\,D^{ |A\cap B|-|B|}
\end{equation}
Let $\rho^a = \bigotimes_{l\in L}\ket{a_l}\bra{a_l}$ and $\rho^b=\bigotimes_{l\in L}\ket{b_l}\bra{b_l}$ be the density matrices corresponding to $\{a_l\}$ and $\{b_l\}$ states respectively. The trace $\text{Tr}_{L}[\rho^a\rho^bS_{A}^i]$ can be rewritten using the swap trick, factoring the link Hilbert space into links that are completely contained in the set $A$ and links that are not. We use $L_A$ to denote the subset of links $L$ that is completely contained in the region $A$ (i.e. both endpoints are in $A$), and the reduced density matrix is defined as $\rho^a_{A} \coloneqq \text{Tr}_{L-L_A}[\rho^a]$ where we are taking the partial trace over links not completely in $A$. After using the swap trick, the fluctuation becomes
\begin{equation}
    \avg{|C_{ab}|^2}=\tilde{C}\sum_{A\in2^{N}}\text{Tr}_{L_A}[\rho_A^a\rho_A^b]\,D^{ |A\cap B|-|B|}
\end{equation}


\bibliographystyle{JHEP}
\bibliography{biblio}


\end{document}